\documentclass[fleqn,usenatbib]{mnras}
\usepackage[T1]{fontenc}
\usepackage{ae,aecompl}
\usepackage{newtxtext,newtxmath}
\usepackage{graphicx}	% Including figure files
\usepackage{amsmath}	% Advanced maths commands
\usepackage{amssymb}	% Extra maths symbols

\def\pow#1#2{#1$\times$10$^{#2}$}

\def\rcm{cm$^{-1}$}

\def\ccm{cm$^{-3}$}

\def\hh{H$_2$} 
 
\def\nhh{$n$(H$_2$)}
\def\tkin{$T_{\rm kin}$}

\title[The interstellar thread of phosphorus between star-forming regions and comets]
{ALMA and ROSINA detections of phosphorus-bearing molecules: the interstellar thread between star-forming regions and comets}
\author[V. M. Rivilla et al.]
{V. M. Rivilla$^{1}$\thanks{E-mail: rivilla@arcetri.astro.it}, %\altaffilmark{2},,
M. N. Drozdovskaya$^{2}$,
K. Altwegg$^{3}$,
P. Caselli$^{4}$,
M. T. Beltr\'an$^{1}$,
\newauthor{F. Fontani$^{1}$,
F.F.S. van der Tak$^{5,6}$,
R. Cesaroni$^{1}$, 
A. Vasyunin$^{7,8}$,
M. Rubin$^{2}$,
F. Lique$^{9}$,}
\newauthor{S. Marinakis$^{10,11}$,
L. Testi$^{1,12,13}$,
and the ROSINA team$^{14}$ 
}
\\
The full list of affiliations appears at the end of the paper in Appendix \ref{affiliations}. \\
}

\date{Accepted 2019 November 21. Received 2019 October 29; in original form 2019 August 8.}

\pubyear{2019}

\hypersetup{draft}

\begin{document}
\label{firstpage}
\pagerange{\pageref{firstpage}--\pageref{lastpage}}
\maketitle

% Abstract of the paper
\begin{abstract}
To understand how Phosphorus-bearing molecules are formed in star-forming regions, we have analysed ALMA observations of PN and PO towards the massive star-forming region AFGL 5142, combined with a new analysis of the data of the comet 67P/Churyumov-Gerasimenko taken with the ROSINA instrument onboard {\it Rosetta}. 
The ALMA maps show that the emission of PN and PO arises from several spots associated with low-velocity gas with narrow linewidths in the cavity walls of a bipolar outflow. %LTE and non-LTE analyses show that 
PO is  more abundant than PN in most of the spots, with the PO/PN ratio increasing as a function of the distance to the protostar. 
%PO and PN are tracing gas at velocities shifted by $\sim$0.5 km s$^{-1}$, suggesting that they come from different layers of the cavity walls. 
Our data favor a formation scenario in which shocks sputter phosphorus from the surface of dust grains, and gas-phase photochemistry induced by UV photons from the protostar allows efficient formation of the two species in the cavity walls. Our analysis of the ROSINA data has revealed that PO is the main carrier of P in the comet, with PO/PN$>$10. Since comets may have delivered a significant amount of prebiotic material to the early Earth, this finding suggests that PO could contribute significantly to the phosphorus reservoir during the dawn of our planet. There is evidence that PO was already in the cometary ices prior to the birth of the Sun, so the chemical budget of the comet might be inherited from the natal environment of the Solar System, which is thought to be a stellar cluster including also massive stars. 
%This hypothesis is supported by the analysis of the P-bearing molecules presented in this work, which has shown chemical similarities between the comet and the AFGL 5142 massive star-forming region.

\end{abstract}


%%%%%%%%%%%%%%%%%%%%%%%%%%%%%%%%%%%%%%%%%%%%%%%%%%

%%%%%%%%%%%%%%%%% BODY OF PAPER %%%%%%%%%%%%%%%%%%

\section{Introduction}

Phosphorus (P) is a key chemical biogenic element for the development of life (\citealt{gulick1955}, \citealt{macia1997}, \citealt{schwartz2006}, \citealt{gamoke2009}, \citealt{fernandez-garcia2017}). While the cosmic abundance of P in the Universe is low relative to hydrogen, P/H $\sim$3$\times$10$^{-7}$ (\citealt{grevesse1998}, \citealt{asplund2009}), its abundance in living organisms is several orders of magnitude higher, e.g., P/H $\sim$10$^{-3}$ in bacteria (e.g., \citealt{fagerbakke1996}). P-compounds are unique in forming large biomolecules, thanks to their extreme structural stability and functional reactivity. Phosphorus is one of the crucial components of deoxyribonucleic acid (DNA) and ribonucleic acid (RNA), phospholipids (the structural components of cellular membranes) and the adenosine triphosphate (ATP) molecule, which stores and transports chemical energy within cells (see, e.g., \citealt{pasek2005}).  Moreover, P-compounds have been proposed as key catalysts and chemical buffers for the formation of nucleotides (\citealt{powder2009}).
%P is therefore essential to life on Earth and can consequently play an important role in Exoplanets (Schaefer & Fegley 2011). 
All of this makes P one of the major limiting nutrient for the development of life (\citealt{redfield1958}), and may make P-bearing molecules important biomarkers in exoplanets (\citealt{sousa-silva2019}).
A fundamental question is how the reservoir of P became biologically available on planets, and in particular on the early Earth. 
%P is in remarkably short supply on Earth, which points towards an extraterrestrial origin. 
In this regard, two research directions can be pursued from the astrophysical/astrochemical point of view. On the one hand, one can study the chemical composition of the interstellar medium (ISM) of the parental molecular clouds that form new stars and planets. On the other hand, one can investigate the chemical compounds of objects found in our Solar System, such as meteorites and comets, which may have delivered prebiotic chemicals to our early Earth.
% during the heavy bombardment phase at $\sim$3.8 Ga. 

The chemistry of P in the ISM is poorly understood. P is a relatively heavy element (atomic mass of 31 Da), and is thought to be synthesized in massive stars and injected into the ISM through supernova explosions (\citealt{koo2013}). As pointed out by \citet{macia1997}, the low number of these massive stars may explain the relatively low cosmic abundance of P relative to hydrogen. Moreover, until very recently, it was thought that P is heavily depleted on the surfaces of insterstellar dust grains in the dense and cold interstellar medium by factors of 600$-$10$^4$ (e.g., \citealt{turner1990}, \citealt{wakelam2008}). This would make the detection of P-bearing molecules in the gas phase through the conventional rotational spectroscopy difficult. Indeed, unlike other biogenic elements (C, N, O, S), P is barely detected in the ISM. The ion P$^{+}$ was detected in several diffuse clouds (\citealt{jura1978}), and only a few simple P-bearing species (PN, PO, CP, HCP, C$_2$P, PH$_3$) have been identified towards the circumstellar envelopes of evolved stars (\citealt{guelin1990}; \citealt{agundez2007}; \citealt{tenenbaum2007}, \citealt{milam2008}, \citealt{halfen2008}, \citealt{debeck2013}, \citealt{agundez2014}, \citealt{ziurys2018}). 
%Only PH$_3$ has been detected in the atmospheres of Jupier and Saturn (\citealt{irwin2004,fletcher2009}).
In star-forming regions, only PN was detected before 2016 towards a handful of sources (\citealt{turner1987}; \citealt{ziurys1987}; \citealt{turner1990}; \citealt{caux2011}; \citealt{yamaguchi2011}). In the last years, a considerable step forward has been made.  Since there is growing evidence that out Solar System was born in a warm and massive dense core with high-mass stars (\citealt{adams2010}, \citealt{pfalzner2015}, \citealt{taquet2016}, \citealt{drozdovskaya2018}, \citealt{lichtenberg2019}), several works have been devoted to studying P-bearing molecules in massive cores, whose chemistry may be inherited by future Solar-like systems. The molecule PN has been detected in other massive star-forming regions (\citealt{fontani2016}; \citealt{mininni2018}; \citealt{fontani2019}), and PO has been detected for the first time in two massive star-forming regions, with an abundance ratio of PO/PN in the range of 1.8$-$3  (\citealt{rivilla2016}). Afterwards, new detections of PO in shocked material (\citealt{lefloch2016}; \citealt{rivilla2018}; \citealt{bergner2019}) have confirmed that PO seems to be more abundant than PN in the ISM.

The relatively low number of detections of PN and PO have prevented so far a good understanding of its formation. Three main mechanisms have been proposed: (i) shock-induced formation (\citealt{yamaguchi2011}; \citealt{aota2012}; \citealt{lefloch2016}; \citealt{rivilla2018}; \citealt{mininni2018}); (ii) high-temperature gas-phase chemistry (\citealt{charnley1994}); and (iii) gas-phase formation during the cold collapse phase of the parental core (\citealt{rivilla2016}). To date, this debate has been strongly limited due to the lack of information about the spatial distribution of the emission of P-bearing molecules in star-forming regions. 
Therefore, interferometric maps of P-bearing molecules are needed to discriminate among the different proposed mechanisms.

The knowledge about P in our Solar System is also limited due to the low number of detections.  Phosphine, PH$_3$, has been observed in the atmospheres of Jupiter and Saturn (\citealt{bregman1975}, \citealt{ridgway1976}, \citealt{larson1977}, \citealt{weisstein1994}, \citealt{irwin2004}, \citealt{fletcher2009});  and P has been identified in meteorites in the form of  the mineral schreibersite (\citealt{pasek2005}) and phosphoric acids (\citealt{schwartz2006}). Traces of P may have been present in the dust of comet Halley, but it was not identified in Stardust grains of comet Wild 2 (\citealt{macia2005}). More recently, the in-situ measurements of the {\it Rosetta} mission claimed the presence of P in the comet 67P/Churyumov-Gerasimenko (67P/C-G, hereafter; \citealt{altwegg2016}), although the parent molecule(s) could not be determined.

In this work, we combine the search for P-bearing molecules in the star-forming region AFGL 5142 using the Atacama Large Millimeter/Submillimeter Array (ALMA) with new analysis of the data of the coma of the comet 67P/C-G taken with the {\it Rosetta} Orbiter Spectrometer for Ion and Neutral Analysis (ROSINA) instrument.
This comparison will allow us to establish if the pristine chemical composition of the comet, in particular the P-bearing reservoir, may have been inherited from a parental molecular core similar to the one that formed our Sun.

AFGL 5142 is a star-forming region in the Perseus arm, where low-mass and high-mass star formation is ongoing simultaneously (\citealt{hunter1999}). It is located at a relatively close distance, 2.14 kpc (\citealt{burns2017}), which allows us to study the molecular emission at high spatial resolution. 
Several H$_2$O and CH$_3$OH masers have been identified in the region (\citealt{hunter1995,goddi2006,goddi2007}).
The center of the region harbors a dust millimeter core (MM position, hereafter, see Fig. \ref{fig-bipolar-outflow}) detected by \citet{hunter1999}, which actually consists of five millimeter sources, some of them associated with hot cores with high temperatures in the range of 90$-$250 K  (\citealt{zhang2007}).
\citet{hunter1999} detected a SiO outflow in the northeast-southwest direction powered by this central protocluster. Observations of other molecular species such as CO, HCN, HCO$^+$ (\citealt{zhang2007,liu2016}) confirmed the presence of abundant shocked material. \citet{busquet2011} revealed, based on observations of N$_2$H$^+$, the presence of a cold starless core located $\sim$12$^{\prime\prime}$ to the west (see Fig. \ref{fig-bipolar-outflow}). Several PN transitions have been detected with single-dish observations (\citealt{fontani2016}; \citealt{mininni2018}). 
The presence of varied physical conditions in a single region makes AGFL 5142  a well suitable laboratory to test the different mechanisms proposed for the formation of P-bearing species. It harbours three different environments where each mechanism could be dominant: i) a system of low- and high-mass protostars whose heating produces a chemically-rich molecular hot core (gas-phase chemistry scenario); ii) abundant shocked material produced by outflowing material (\citealt{zhang2002}; \citealt{liu2016}; shock scenario); and iii) a starless cold core (\citealt{busquet2011}; cold collapse scenario). 

%% Hunter99: the lymann continuum flux in the centimeter indicated an exciting source a B2 or earlier zero-age main-sequence star.
%The presence of very different physical conditions in the region, i.e., high-temperature gas in the central dusty core, extended shocked emission associated with outflowing material, and cold gas in the starless core, allows us to test the different formation routes proposed to form P-bearing molecules.

67P/C-G is a Jupiter-family comet with a period of $\sim$6.5 years. Several studies suggested that it experienced a close encounter with Jupiter in February 1959, which reduced its perihelion distance from 2.7 to the current 1.2 au (\citealt{lamy2007}). Its nucleus has a bilobate shape and is approximately 4.3$\times$2.6$\times$2.1 km in size (\citealt{jorda2016}). The {\it Rosetta} mission escorted the comet pre- and post- its August 2015 perihelion passage, revealing that the dusty surface covers an icy interior (\citealt{fornasier2016}). The most recent analysis of mission data suggest that 67P/C-G is indeed a primordial rubble pile containing non-thermally processed materials that were once part of the protoplanetary disk that evolved into our modern-day Solar System (e.g., \citealt{altwegg2015,davidsson2016,alexander2018}).

This work is organised as follows. In Section \ref{observations} we present the ALMA and ROSINA data. We present our results in Section \ref{analysis} and discuss their implications for the formation of P-bearing molecules in star- and planet-forming regions in Section \ref{discussion}. Finally, we summarize our main findings in Section \ref{conclusions}.

\begin{table}
\begin{center}
\caption[]{Molecular transitions studied in this ALMA dataset, from the CDMS molecular database.}
\tabcolsep 2.5pt
\begin{tabular}{c c c c c }
\hline \noalign {\smallskip}
Molecule & Transition & Frequency  &  log$A_{\rm ij}$  &E$_{\rm up}$   \\
                &                 &   (GHz) &   (s$^{-1}$) & (K)  \\
\hline 
PN  & N=2$-$1, J=2$-$2 &  93.9782  &  -5.13724 &  6.8  \\ %
PN  & N=2$-$1, J=1$-$0 &  93.9785  &  -4.79039 & 6.8  \\   %
PN  & N=2$-$1, J=2$-$1 &  93.9798  & -4.66014 & 6.8  \\ %
PN  & N=2$-$1, J=3$-$2 &  93.9799  & -4.53516 &  6.8  \\ %
PN  & N=2$-$1, J=1$-$2 &  93.9808  & -6.09138 &  6.8  \\
PN  & N=2$-$1, J=1$-$1 &  93.9823  & -4.91538 &  6.8  \\  %
\hline
PO &   J=5/2$-$3/2, $\Omega$=1/2, F=3$-$2, l=e & 108.9984 & -4.67162  &  8.4   \\
PO &   J=5/2$-$3/2, $\Omega$=1/2, F=2$-$1, l=e & 109.0454 &-4.71670 &   8.4  \\
PO &   J=5/2$-$3/2, $\Omega$=1/2, F=3$-$2, l=f &109.2062 & -4.69959 &  8.4   \\
PO &   J=5/2$-$3/2, $\Omega$=1/2, F=2$-$1, l=f & 109.2812 & -4.71444 &  8.4   \\
\hline
SO & N=2$-$1, J=3$-$2 & 109.2522 & -4.95802  & 21.1   \\
\hline
%\hline \noalign {\smallskip}
\end{tabular}
\label{table-molecular-transitions}
\end{center}
\end{table}

%%%%%%%%%%%%%%%%%%%%%%%%%%%%%%%%%%%%%
\begin{table}
%\scriptsize
\centering
\caption{Coordinates of the positions in the AFGL 5142 region studied in this work.}
\begin{tabular}{c c c} 
 \hline
Position & RA (J2000) & DEC (J2000)  \\
Position & h : m : s  & $^{\circ}$ : $\arcmin$ : $\arcsec$  \\
 \hline
 P1 & 05:30:48.38  & 33:48:14.5  \\  
 P2   &   05:30:48.18  & 33:47:45.1 \\ 
 P3    &  05:30:47.45  & 33:47:43.1 \\        
 P4    &  05:30:47.09  & 33:47:36.3   \\ 
 P5     &  05:30:49.54  & 33:47:35.7  \\    
 P6     &  05:30:49.78  & 33:47:33.3\\     
 P7     &  05:30:49.86  & 33:47:30.3  \\     
 MM     &  05:30:48.02  & 33:47:54.2  \\
 SC     &  05:30:48.95  & 33:47:52.9 \\     
 HV-b     &  05:30:47.83  & 33:47:47.3  \\  
 HV-r       &   05:30:48.12&  33:48:00.2         \\   
\hline
\end{tabular}
\label{table-positions}
\end{table}

%%%%%%%%%%%%%%%%%%%%%%%%%%%%%%%%%%%%%
\begin{table*}
%\scriptsize
%%%%%%%%%%%%%% google Spreedsheet AFGL5142 - ALMA DATA
\centering

\caption{Results of the LTE analysis of the PN, PO and SO transitions detected towards the different AFGL 5142 positions studied in this work. Some of the positions exhibit several velocity components (different rows in this table).}
\tabcolsep 1.75pt
\begin{tabular}{c c c c c c c c c c c c c c c c} 
    \hline
Region   		   & \multicolumn{4}{c}{PN} & & \multicolumn{4}{c}{PO} & & \multicolumn{4}{c}{SO}  \\  \cline{2-5}  \cline{7-10}  \cline{12-15}
 & 	   $N$  & $v_{\rm LSR}$ & $\Delta v$  & $\tau^{(a)}$ & & $N$  & $v_{\rm LSR}$ & $\Delta v$ & $\tau^{(a)}$ & & $N$  & $v_{\rm LSR}$ & $\Delta v$  & $\tau^{(a)}$   \\ 

      & ($\times$10$^{12}$ cm$^{-2}$)  & (km s$^{-1}$) & (km s$^{-1}$)  & & & ($\times$10$^{12}$ cm$^{-2}$)  & (km s$^{-1}$) & (km s$^{-1}$) & & &($\times$10$^{14}$ cm$^{-2}$)  & (km s$^{-1}$) & (km s$^{-1}$)  &  \\  
		\hline
		P1     &   1.8$\pm$0.2 & -0.8$\pm$0.2 & 2.6$\pm$0.3 & 0.047$\pm$0.006 & & < 2.0  & -0.8  & 2.6  & - &  & 3.8$\pm$0.2  & -1.53$\pm$0.04 &  3.8$\pm$0.2   & 0.048$\pm$0.001 \\
  		   & < 0.3 & 2.1 & 7.1 & -  &  & < 3.3   & 2.1  & 7.1  & -  &   & 1.84$\pm$0.08  & 2.1$\pm$0.2 &  7.1$\pm$0.4    & 0.012$\pm$0.001 \\
              
        \hline
		P2    & 3.1$\pm$0.2 & -2.1$\pm$0.2 & 5.0$\pm$0.4 &  0.041$\pm$0.004  & & 1.7$\pm$0.6 & -2.1 &  5.0   &  0.002$\pm$0.001  & & 5.25$\pm$0.06 & -2.1$\pm$0.1 & 5.8$\pm$0.1   & 0.043$\pm$0.001\\ 
        	      & < 0.4 & -1.97 & 2.0 & -  &     & < 0.1 & -1.97 &  2.0  & -  & & 1.78$\pm$0.05 & -1.97$\pm$0.03 & 2.05$\pm$0.07   &  0.042$\pm$0.001\\ 
               & < 0.2 & -3.76 & 1.0 & -  &     & < 0.1 &  -3.76  &  1.0  &- & & 0.68$\pm$0.03 & -3.76$\pm$0.03 & 1.05$\pm$0.06    &  0.031$\pm$0.002\\ 
               & < 0.2 & -6.2 & 1.7 & -  &       & < 0.1 &   -6.2   & 1.7  & -& & 0.40$\pm$0.02 & -6.2$\pm$0.05 & 1.7$\pm$0.1  &  0.011$\pm$0.001\\ 
        \hline    
		P3    & 2.3$\pm$0.2 & -3.47$\pm$0.04 & 1.69$\pm$0.09 &  0.092$\pm$0.006  &&  3.1$\pm$0.3 &  -3.04$\pm$0.06 & 1.4$\pm$0.2  &  0.012$\pm$0.002   & & 3.5$\pm$0.2 &  -3.21$\pm$0.04 & 1.69$\pm$0.09   & 0.099$\pm$0.007 \\        
        \hline        
		P4     & 5.5$\pm$0.2  & -5.14$\pm$0.03 & 1.50$\pm$0.05 & 0.24$\pm$0.01   && 8.8$\pm$0.6  & -4.68$\pm$0.04  & 1.2$\pm$0.1   & 0.040$\pm$0.006    & & 5.8$\pm$0.2 & -4.66$\pm$0.03 & 1.29$\pm$0.06    & 0.22$\pm$0.02 \\ 
        \hline     
  	P5     & 2.4$\pm$0.2 & -3.49$\pm$0.04  & 1.16$\pm$0.09  & 0.14$\pm$0.02 & & 6.2$\pm$0.8  & -3.25$\pm$0.05 & 1.1$\pm$0.2 &  0.031$\pm$0.006 &  & 1.71$\pm$0.09 & -3.12$\pm$0.03 & 0.92$\pm$0.06    & 0.088$\pm$0.007 \\    
   	P6     & 2.2$\pm$0.2 & -3.26$\pm$0.05 & 1.3$\pm$0.2 & 0.12$\pm$0.02  & & 4.8$\pm$0.8 & -2.77$\pm$0.06 & 0.9$\pm$0.2 & 0.031$\pm$0.006   & & 1.6$\pm$0.2 & -2.81$\pm$0.04 & 1.0$\pm$0.1   & 0.08$\pm$0.01\\     
    P7    & 1.3$\pm$0.3 & -3.3$\pm$0.2 & 1.7$\pm$0.3 & 0.11$\pm$0.03   &  &  < 1.2 & - & - & -  &  & 1.10$\pm$0.06 & -3.03$\pm$0.03 & 0.94$\pm$0.07    & 0.055$\pm$0.004\\     
     \hline       
MM      & 0.8$\pm$0.1  & -3.9$\pm$0.3 & 3.9$\pm$0.7 & 0.031$\pm$0.007  & & < 0.6 & -3.9 & 3.9  &  -   &  & 28.0$\pm$0.6 & -2.85$\pm$0.05 & 4.7$\pm$0.1   & 0.278$\pm$0.009 \\
\hline
%		SC     & < & - & - &   & & <  & - & - &    & & & &    &\\            
%        \hline  
 	HV-b  &   < 0.18 & -4.1 & 2.1  & -  && < 0.4 & -4.1 & 2.1 &  - &  &  1.92$\pm$0.08   & -4.07$\pm$0.05 & 2.1$\pm$0.1  &  0.044$\pm$0.003 \\  
     	                & < 0.35  & -7.9 & 6.1 & -  &  & < 0.6 & -7.9 & 6.1 & -  & & 4.62$\pm$0.08  & -7.95$\pm$0.05 &  6.1$\pm$0.2    & 0.035$\pm$0.001\\ 
             \hline
    HV-r              & < 0.16   & -3.96 & 4.85 & -  &  & < 0.5 & -3.96 & 4.85 & -  & & 3.21$\pm$0.05  & -3.96$\pm$0.04 &  4.85$\pm$0.09   & 0.031$\pm$0.001\\ 
                           & < 0.28  & -4.22 & 6.2 &  - &  & < 0.3 & -4.22 & 6.2 &  - & & 2.78$\pm$0.07  & 4.22$\pm$0.07 &  6.2$\pm$0.2  &  0.021$\pm$0.001\\
                           & < 0.32   & -0.21 & 1.5 & -  &  & < 0.6 & -0.21 & 1.5 & -  & & 0.32$\pm$0.04  & 0.21$\pm$0.05 &  1.5$\pm$0.2  & 0.010$\pm$0.002 \\ 
        \hline
	\end{tabular}
\label{table-sample}
$^{(a)}$ Opacity of the hyperfine transition with highest  $\tau$.
\end{table*}

%%%%%%%%%%%%%%%%%%%%%%%%%%%%%%%%%%%%%
\begin{table}
\centering
\caption{Molecular abundance ratios at the AFGL 5142 positions and in the comet 67P/C-G. Different rows in a single position correspond to the different velocity components shown in Table \ref{table-sample}.}
\begin{tabular}{c c c c } 
    \hline
   	& PO/PN &   SO/PN & SO/PO  \\ 
		\hline
		 \multicolumn{4}{c}{AFGL 5142 positions} \\
		 \hline
		P1     &   < 1.1   & 207$\pm$9 & $> $195\\
  		     &    -   &  $>$ 650 & $>$57 \\    
  		     \hline
		P2   &   0.6$\pm$0.2 & 170$\pm$11 & 309$\pm$107\\ 
        	  &     - &  $>$ 470 & $>$1573  \\ 
                &  - &   $>$ 340  &  $>$1202 \\ 
                  &   - &    $>$ 200   &   $>$707\\ 
     \hline
		P3    &  1.4$\pm$0.3   & 151$\pm$14 &  112$\pm$17\\   
		  		     \hline    
		P4    & 1.6$\pm$0.2  & 107$\pm$7  & 66$\pm$7 \\
		  		     \hline
  	P5     &   2.6$\pm$0.6  &  72$\pm$9 & 27$\pm$5  \\    
  	  		     \hline
   	P6     &   2.2$\pm$0.6  & 70$\pm$12 & 33$\pm$10\\   
   	  		     \hline
    P7     &   < 0.9  & 84$\pm$18 & $>$90 \\        
      		     \hline  
MM   &   < 0.8    &  3370$\pm$580 & $>$4498 \\
  		     \hline
%		SC     &   - & -  \\       
%		  		     \hline      
 	HV-b     &  -  &  $>$ 1057 & $>$522 \\  
     	     &     -  &  1303 &  $>$ 778 \\ 
     	       		     \hline
    HV-r       &      -&  $>$ 2042  & $>$630\\ 
               &       - & $>$ 979 & $>$983\\
               &      -       &  $>$ 101 & $>$57\\ 
%                 		     \hline
%    HV-r       &      -&  $>$ 2042  & $>$181\\       
    \hline           		     
         		 \multicolumn{4}{c}{comet 67P/Churyumov-Gerasimenko} \\   
         		 \hline    
      &  $>$10    &  $>$60  & $\sim$6 \\           
         		  		     
        \hline
\end{tabular}
\label{table-ratios}
\end{table}

\section{Observations}
\label{observations}

\subsection{ALMA observations}
\label{observations-ALMA}

We carried out interferometric observations using 40 antennas of the Atacama Large Millimeter/Submillimeter Array (ALMA) in Cycle 4 between January and April of 2017 as part of the project 2016.1.01071.S (PI: Rivilla). 
The phase center was RA= 5h 30m 48.0s,  DEC=33$^{\circ}$ 47$^{\prime}$ 54.0$^{\prime\prime}$.
The observations were performed in Band 3 with the array in the C40$-$3 configuration with baselines ranging from 15 to 460 m. The digital correlator was configured in 12 different spectral windows (spw's) with channel widths of 122 kHz ($\sim$0.35 km s$^{-1}$), which cover the PN(2$-$1) transition, the J=5/2$-$3/2, $\Omega$=1/2 quadruplet of PO, and the SO 2$_{3}-$1$_{2}$ transition (see Table \ref{table-molecular-transitions}). The precipitable water vapor (pwv) during the observations was in the range of 1.5$-$5.6 mm. Flux and bandpass calibration were obtained through observations of J0510+1800. The phase was calibrated by observing J0547+2721 and J0552+3754. The on-source observing time was about 2.5 hr. 
The synthesized beams are 2.5$\arcsec\times$1.8$\arcsec$ for the PN map, and  2.1$\arcsec\times$1.6$\arcsec$ for the PO and SO maps. 
The root mean square ({\it{rms}}) of the noise of the maps is $\sim$2 mJy beam$^{-1}$ per channel.  
The data were calibrated and imaged using standard ALMA calibration scripts of the Common Astronomy Software Applications package (CASA)\footnote{https://casa.nrao.edu}. Further analysis was done with MADCUBA\footnote{Madrid Data Cube Analysis on ImageJ is a software developed in the Center of Astrobiology (Madrid, INTA-CSIC) to visualize and analyze astronomical single spectra and datacubes. MADCUBA is available at http://cab.inta-csic.es/madcuba/Portada.html} software package (\citealt{martin2019}).

\subsection{ROSINA measurements}
\label{observations-ROSINA}

The {\it Rosetta} spacecraft of the European Space Agency (\citealt{glassmeier2007}) accompanied the comet 67P/C-G during two years from August 2014 through September 2016. On-board {\it Rosetta} was the instrument suite {\it Rosetta} Orbiter Sensor for Ion and Neutral Analysis (ROSINA) with two mass spectrometers and a pressure sensor (\citealt{balsiger2007}). During the two weeks on October 2014, at a distance of 3 au from the Sun,  but only 10 km from the nucleus, the densities measured in the coma by ROSINA DFMS (Double Focusing Mass Spectrometer) were relatively high allowing the detection of low-abundance species. 
DFMS measures one integer mass with a resolution of $\sim$9000 at full width half maximum (FWHM) at mass 28 Da at a time. The integration time is 20 s. The detector has a one-dimensional array of 512 pixels. A peak can be well approximated by a double Gaussian, whereby the second Gaussian has a $\sim$3 times larger width and a 10 times lower height. For small peaks, one Gaussian is precise enough for most purposes. All peaks on one integer mass have the same widths for the two Gaussians. Details on the data analysis can be found in \citet{leRoy2015} and \citet{calmonte2016}.

%--------------------------------------------------------------------
\section{Analysis and Results}
\label{analysis}

\begin{figure*}
\centering
%%%%%%%% spreedsheet figura1-paper
%\includegraphics[width=15cm]{bipolar-outflow.ps}
%
%\includegraphics[width=18cm]{figura1_paper.eps}
\includegraphics[width=18cm]{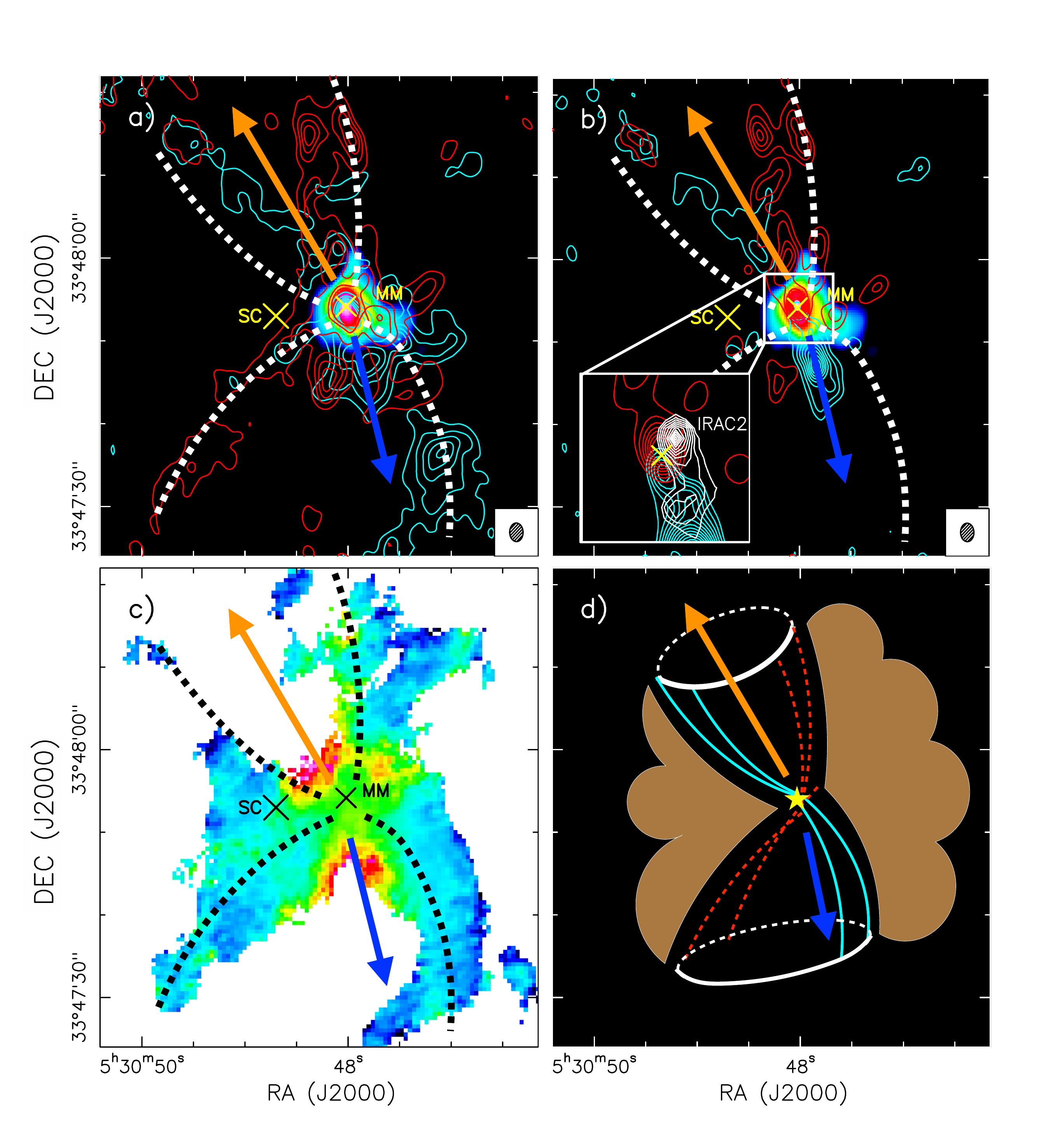}
\hskip10mm
%\includegraphics[width=8.675cm]{map-NH3.eps}
%includegraphics[width=7.5cm]{map-sketch.eps}
 \caption{a) ALMA maps of the SO(2$_{3}-$1$_2$) transition. We show two different velocity ranges:  [-12,-4] km s$^{-1}$ (light blue contours) and [-3.5, 6] km s$^{-1}$ (red contours). The contour levels start at 40/50 mJy km s$^{-1}$ beam$^{-1}$ and increase in steps of 75/100  mJy km s$^{-1}$ beam$^{-1}$ for the blushifted/redshifted emission. The continuum map is shown in color scale, starting from blue ($\sim$2 mJy beam$^{-1}$) to magenta (30 mJy beam$^{-1}$). The beam of the continuum map (2.27$\arcsec\times$1.65$\arcsec$) is indicated in the lower right corner. The cavity of the bipolar outflow is indicated with dashed white curves, and the direction of the redshifted/blushifted lobes of the bipolar outflow are indicated with orange/blue arrows, respectively. The positions of the MM core and the starless core (SC) are indicated with yellow crosses.  b) Same as in panel a), but showing only the high velocity gas:  [0.5,6] km s$^{-1}$ (red) and [-12,-7.5] km s$^{-1}$ (light blue). The contour levels start at 20/50 mJy km s$^{-1}$ beam$^{-1}$ and increase in steps of 25/50  mJy km s$^{-1}$ beam$^{-1}$ for the blushifted/redshifted emission. The inset in the lower left shows a zoom-in of the inner region. In the inset, the contour levels start at 30/100 mJy km s$^{-1}$ beam$^{-1}$ and increase in steps of 10/50  mJy km s$^{-1}$ beam$^{-1}$ for the blushifted/redshifted emission. The emission at 4.5 $\mu$m from Spitzer$-$IRAC2 is overplotted in white contours. c) Ratio between the emission of NH$_3$ (J,K) = (3,3) to (J,K) = (1,1), adapted from \citet{zhang2002} (see this work for futher details). The color scale is a proxy of the gas temperature: red is hotter and dark blue is cooler. The gas is heated at the base of the two molecular outflow cavities. d) Schematic picture of the region. The bipolar outflow is close to the plane of the sky, and it is excavating two cavities in the natal molecular core. }
    \label{fig-bipolar-outflow}
\end{figure*}

\begin{figure*}
\centering
% SCRIPTS: @map-PN-rgb.greg
\includegraphics[width=18cm]{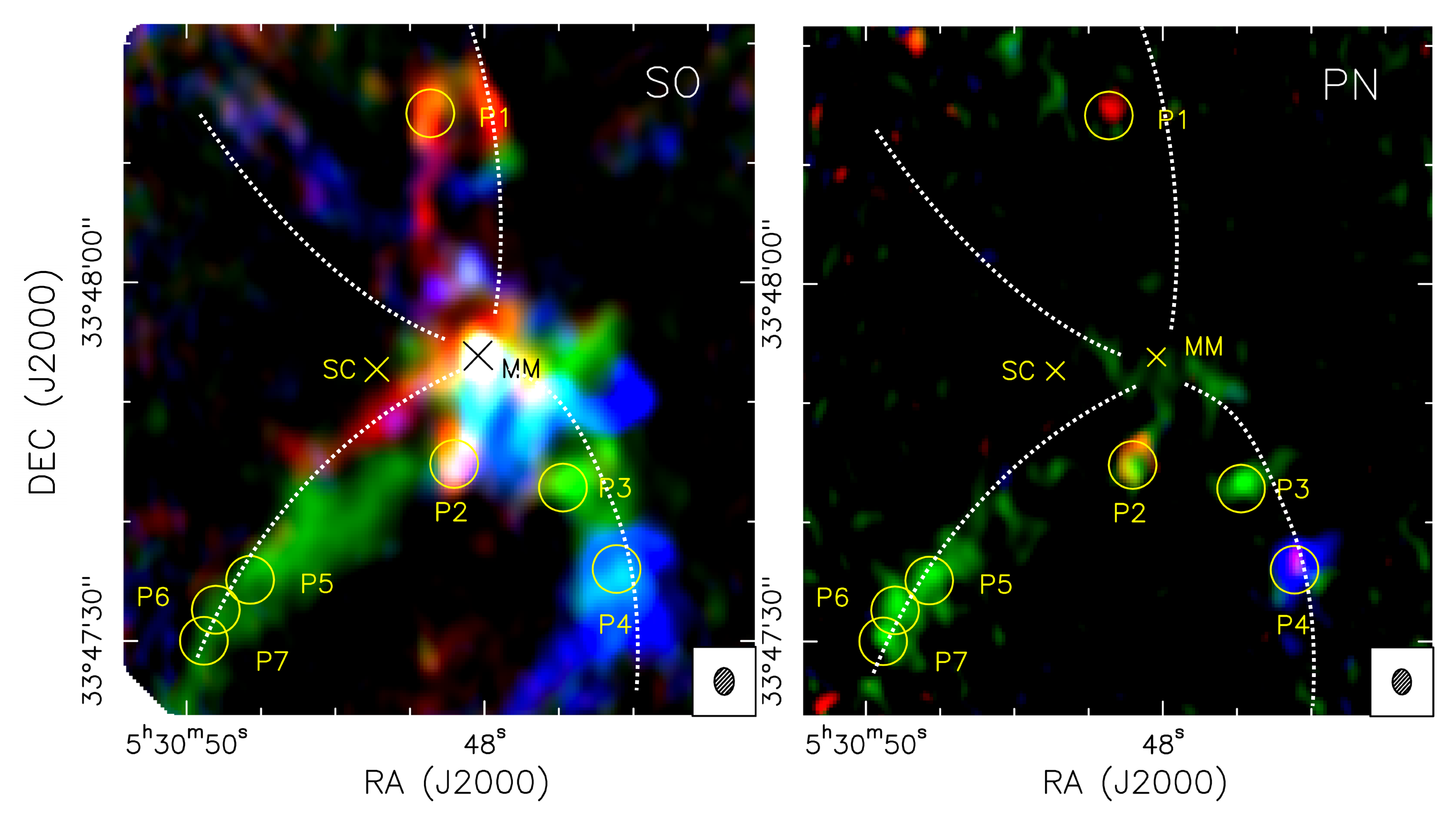}
 \caption{ALMA integrated maps of SO(2$_{3}-$1$_2$)  (left) and PN(2$-$1) (right) in three velocity ranges  [-6.5,-4.5] km s$^{-1}$ (blue); [-4.5,-2.0] km s$^{-1}$ (green); [-2.0, 0] km s$^{-1}$ (red). The cavity of the bipolar outflow is indicated with dashed white curves. The positions of the starless core (SC) and the MM core are indicated with yellow (or black) crosses. The different P-spots (P1$-$P7) are identified with yellow circles. The beams of the observations are indicated in the lower right corner of each panel.}
\label{fig-3colors}
\end{figure*}

\begin{figure}
%\centering
% SCRIPTS: @map-PN-rgb+SO-hv.greg; edit with speedsheet rgb-PN-SOhv
\includegraphics[width=8.5cm]{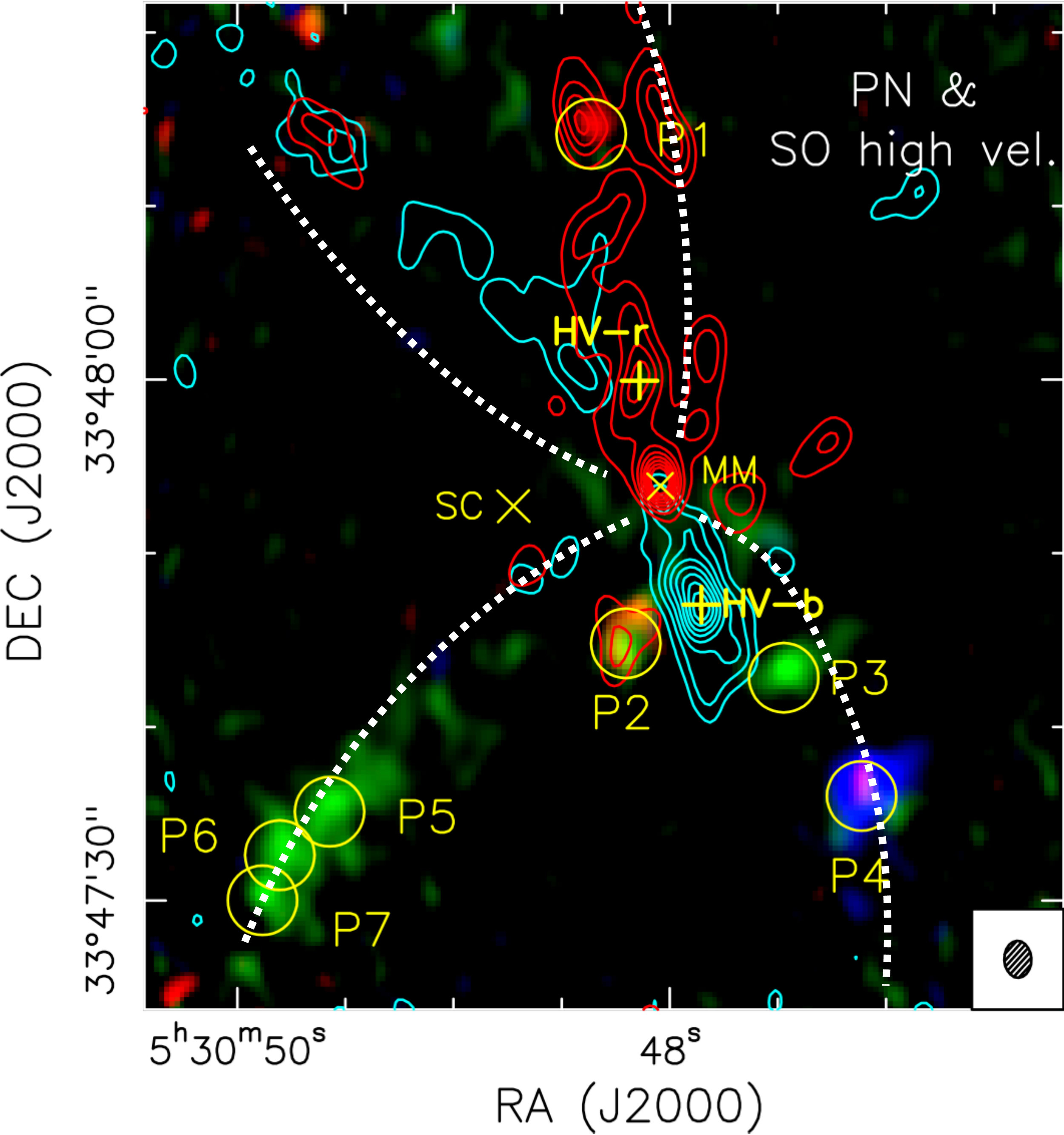}
\caption{ALMA integrated PN(2$-$1) emission map (in colours) as shown in the right panel of Fig. \ref{fig-3colors}, with the high velocity SO emission (in contours) shown in Fig. \ref{fig-bipolar-outflow}b overplotted. The positions of the starless core (SC) and the MM core are indicated with yellow crosses. The different P-spots (P1$-$P7) are identified with yellow circles. The positions of the high-velocity spots HV-r and HV-b are indicated with yellow plus signs. The cavity of the bipolar outflow is indicated with dashed white curves.}
\label{fig-PN-SO-HV}
\end{figure}

\begin{figure*}
\centering
%  SCRIPTS: @plot-spectra.greg, @plot-spectra2.greg, @plot-spectra3.greg
\includegraphics[width=12.5cm]{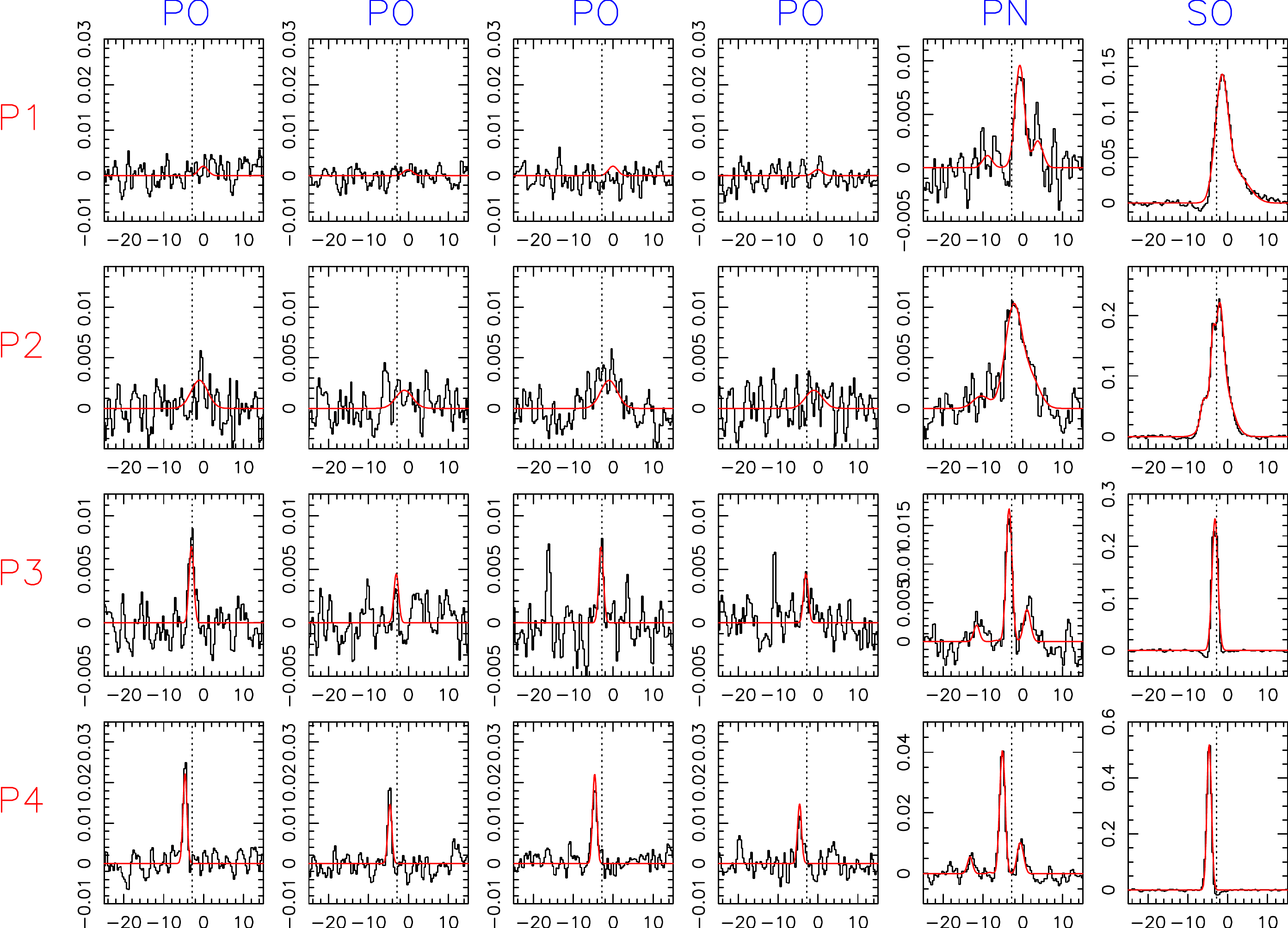}
\vskip0.1mm
\hskip-1mm
\includegraphics[width=12.5cm]{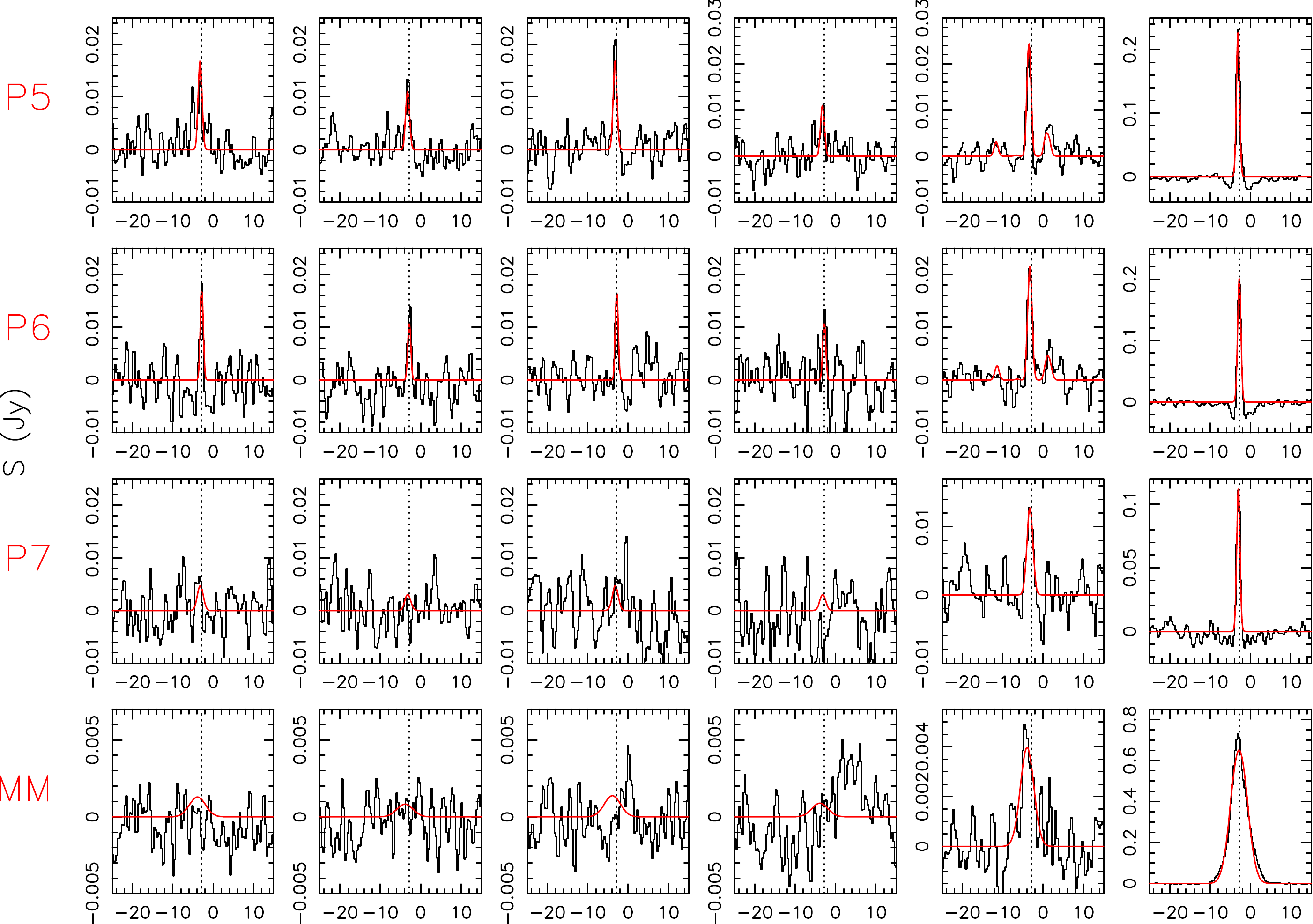}
\vskip1mm
\hskip-3.5mm
\includegraphics[width=12.6cm]{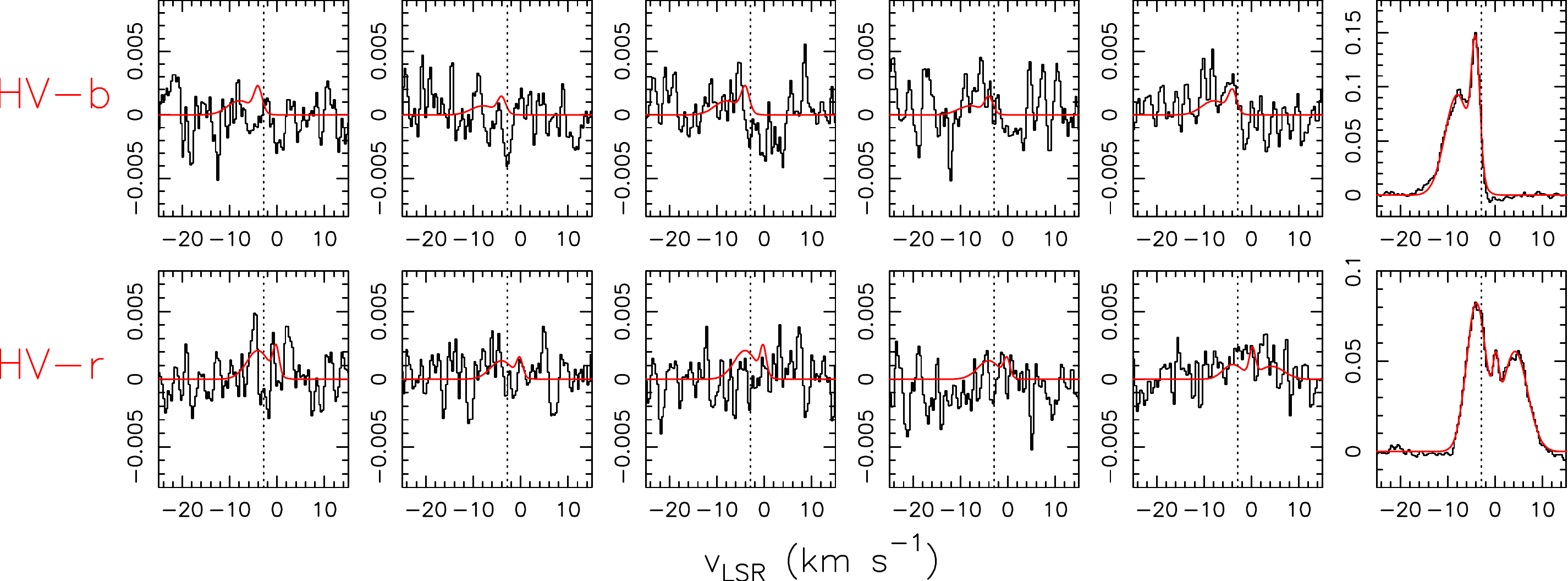}
 \caption{ALMA spectra of PO, PN and SO (molecular transitions listed in Table \ref{table-molecular-transitions}) towards different positions of the AFGL 5142 star-forming region, indicated to the left of the panels. The dashed vertical lines indicate the systemic velocity of the central core, -2.85 km s$^{-1}$. The red lines are LTE fits to the data obtained as explained in the text.}
    \label{fig-spectra}
\end{figure*}

\begin{figure*}
\centering
% SCRIPTS:   @map-PN-PO_P5.greg,  @map-PN-PO.greg, @map-PN-PO_P567.greg
\hskip-8mm
\includegraphics[width=9cm]{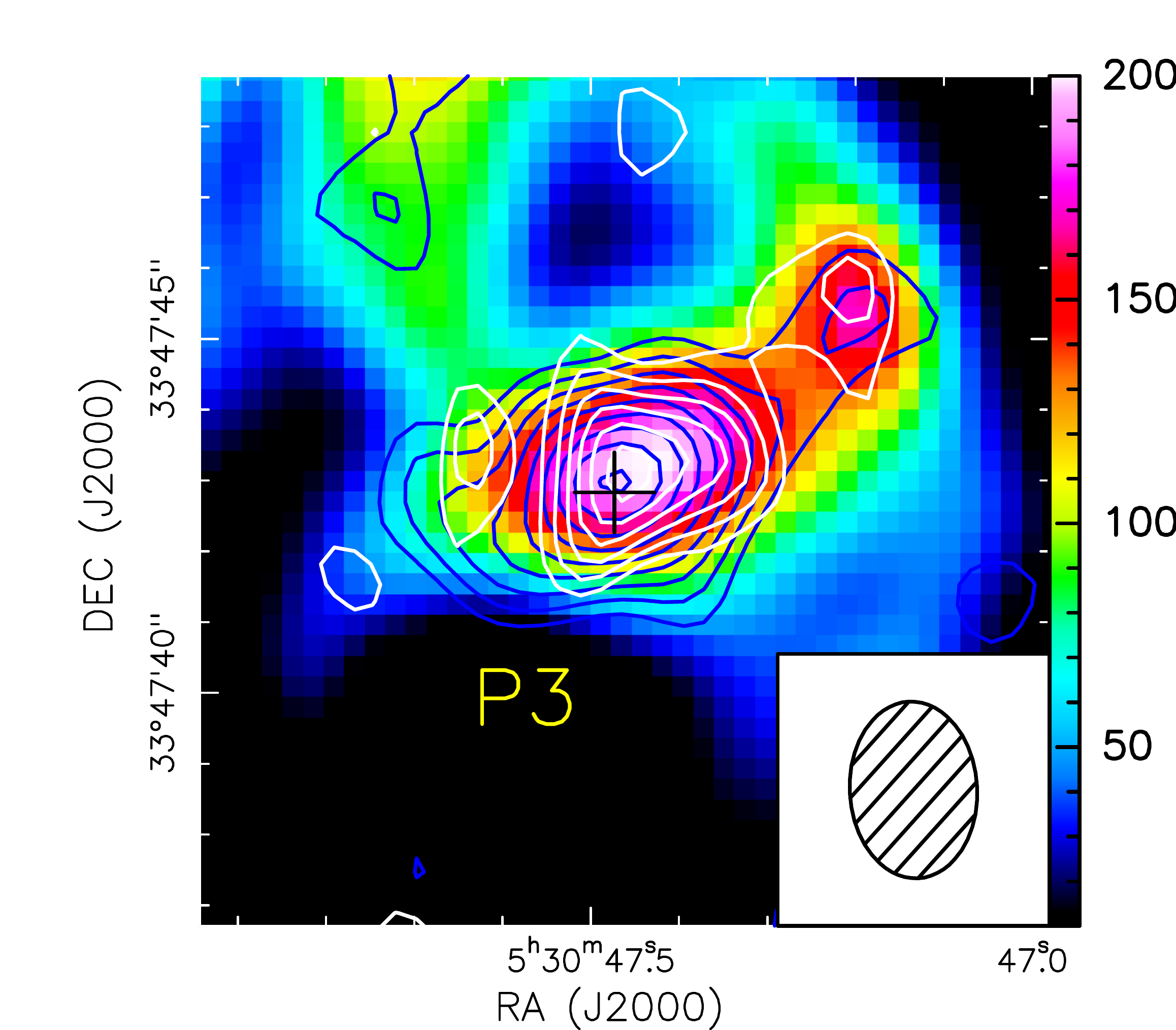}
\hskip-2mm
\includegraphics[width=9cm]{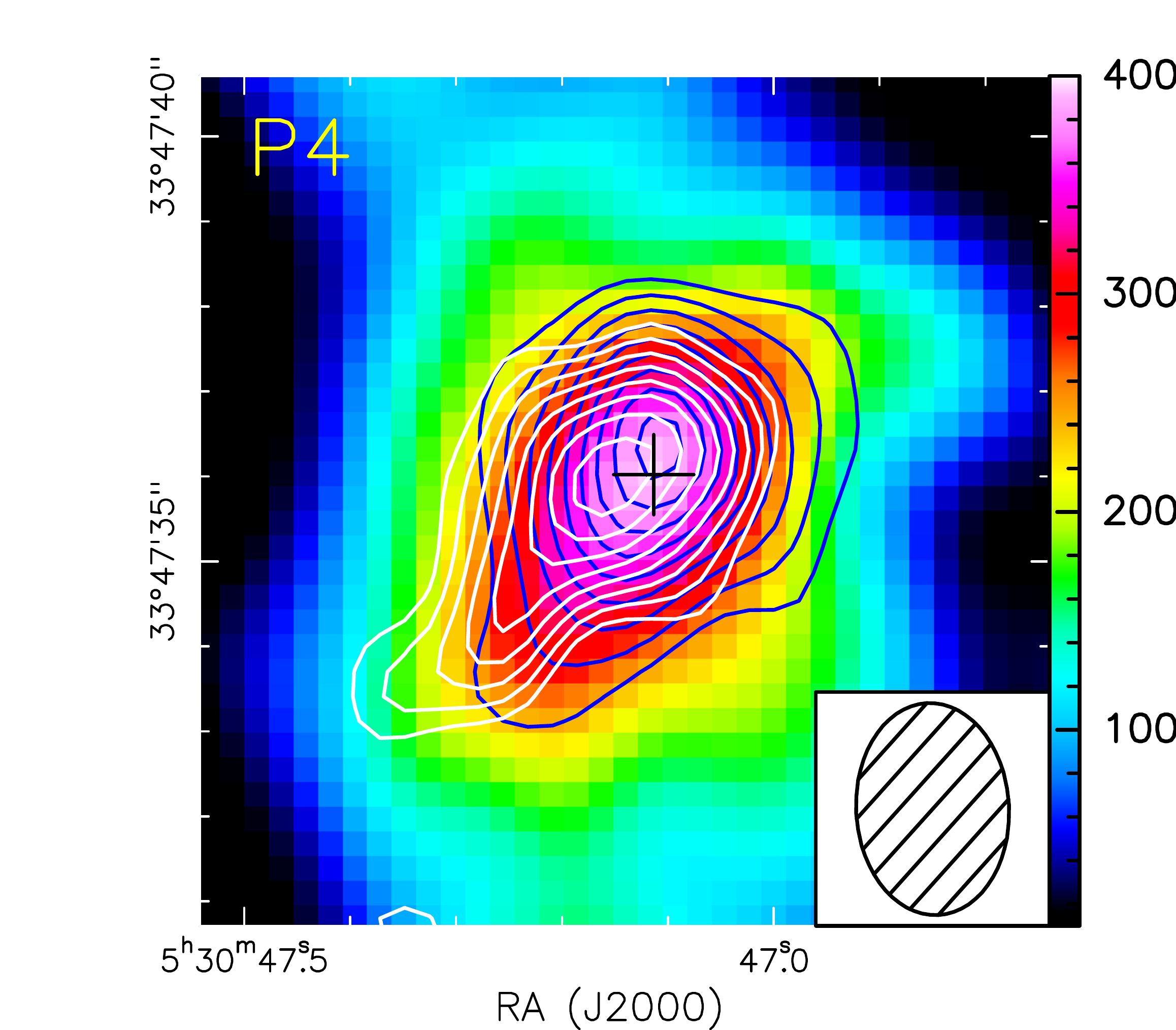}
\vskip5mm
\includegraphics[width=10cm]{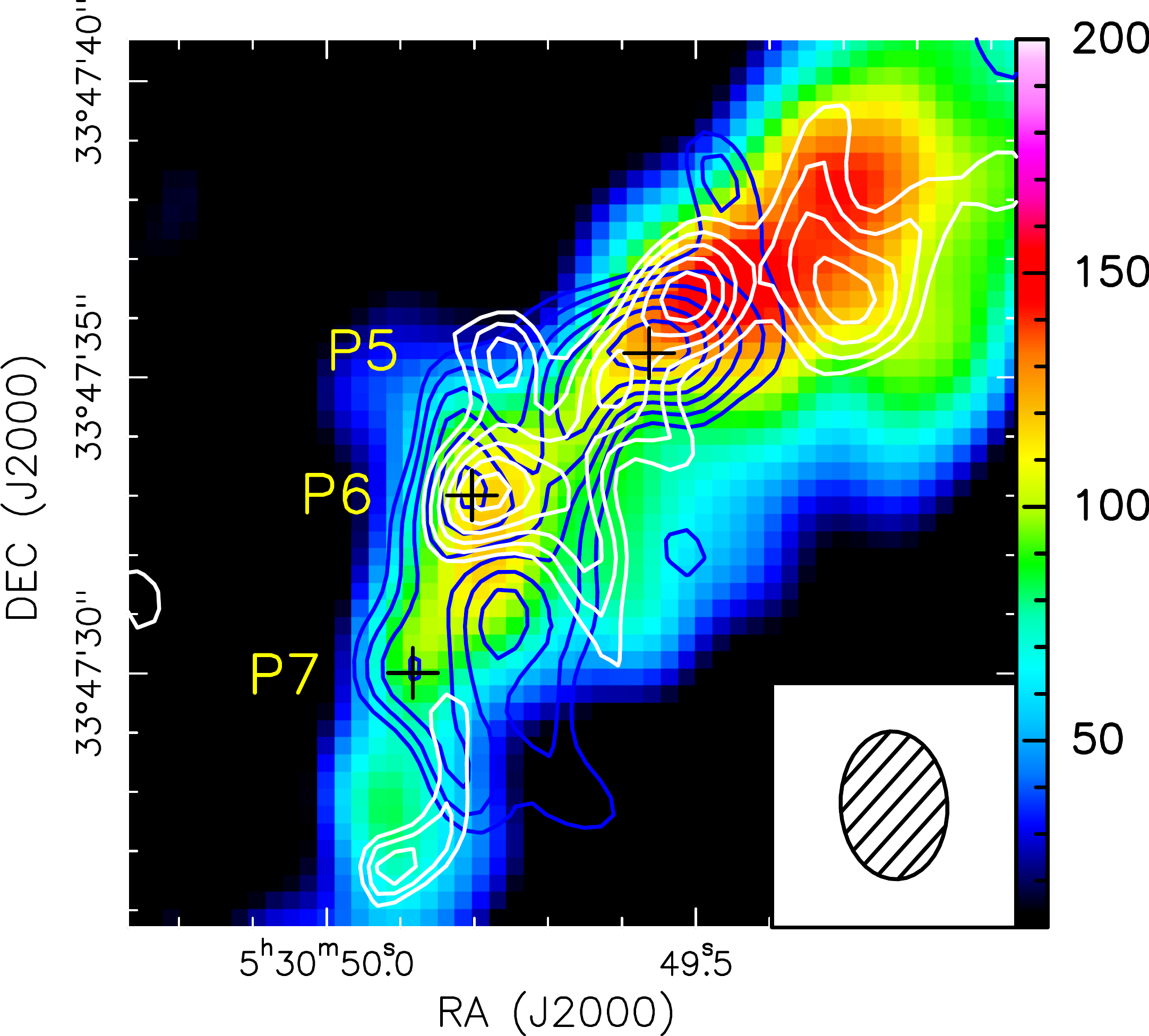}
\caption{Zoom-in view of the P3 spot (upper left panel), P4 spot (upper right panel) and P5, P6 and P7 spots (lower panel). The color scale represents the integrated emission of the SO 2$_{3}-$1$_2$ transition, in mJy beam$^{-1}$ km s$^{-1}$. The blue contours indicate the total integrated emission of the PN(2$-$1) transitions at 93.978209, 93.978477 and 93.97978 GHz. The contour levels start at 20/6/13 mJy km s$^{-1}$ beam$^{-1}$ and increase in steps of 5/3/3 mJy km s$^{-1}$ beam$^{-1}$ for the P3/P4/P5-P6-P7 regions, respectively. The white contours correspond to the combined integrated emission of the four hyperfine PO transitions (see Table \ref{table-molecular-transitions}). The contour levels start at 20/6/20 mJy km s$^{-1}$ beam$^{-1}$ and increase in steps of 5/3/5 mJy km s$^{-1}$ beam$^{-1}$ for the P3/P4/P5-P6-P7 regions, respectively. The positions of the different P-spots (Table \ref{table-positions}) are indicated with black plus signs.}
\label{fig-PO}
\end{figure*}

 \begin{figure}
\centering
%  SCRIPT: @plot_vel_vs_vel.greg
\includegraphics[width=7.5cm]{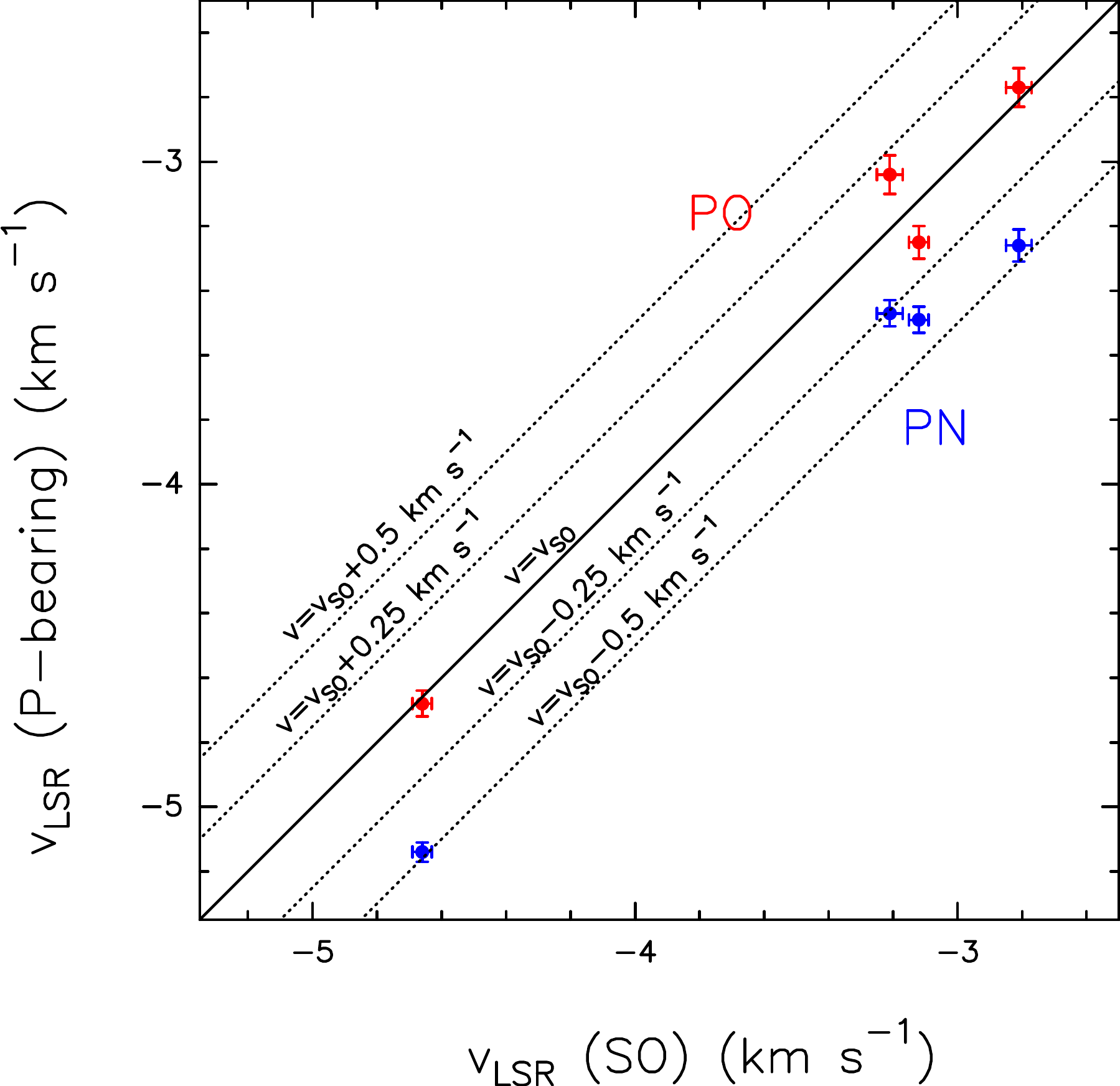}
 \caption{Velocity of PN (blue points) and PO (red points) as a function of SO velocity towards the P-spots for which the three species have been detected and the velocity was left as a free parameter in the AUTOFIT analysis: P3, P4, P5 and P6 spots, located in the southern (blueshifted) outflow cavity. The solid black line denotes equal velocities, and the dashed lines show differences of $\pm$0.25 and $\pm$0.5 km s$^{-1}$.}
    \label{fig_vel_vs_vel}
\end{figure}

\begin{figure}
\centering
%  SCRIPT: @plot_N_vs_vel.greg
\includegraphics[width=7.5cm]{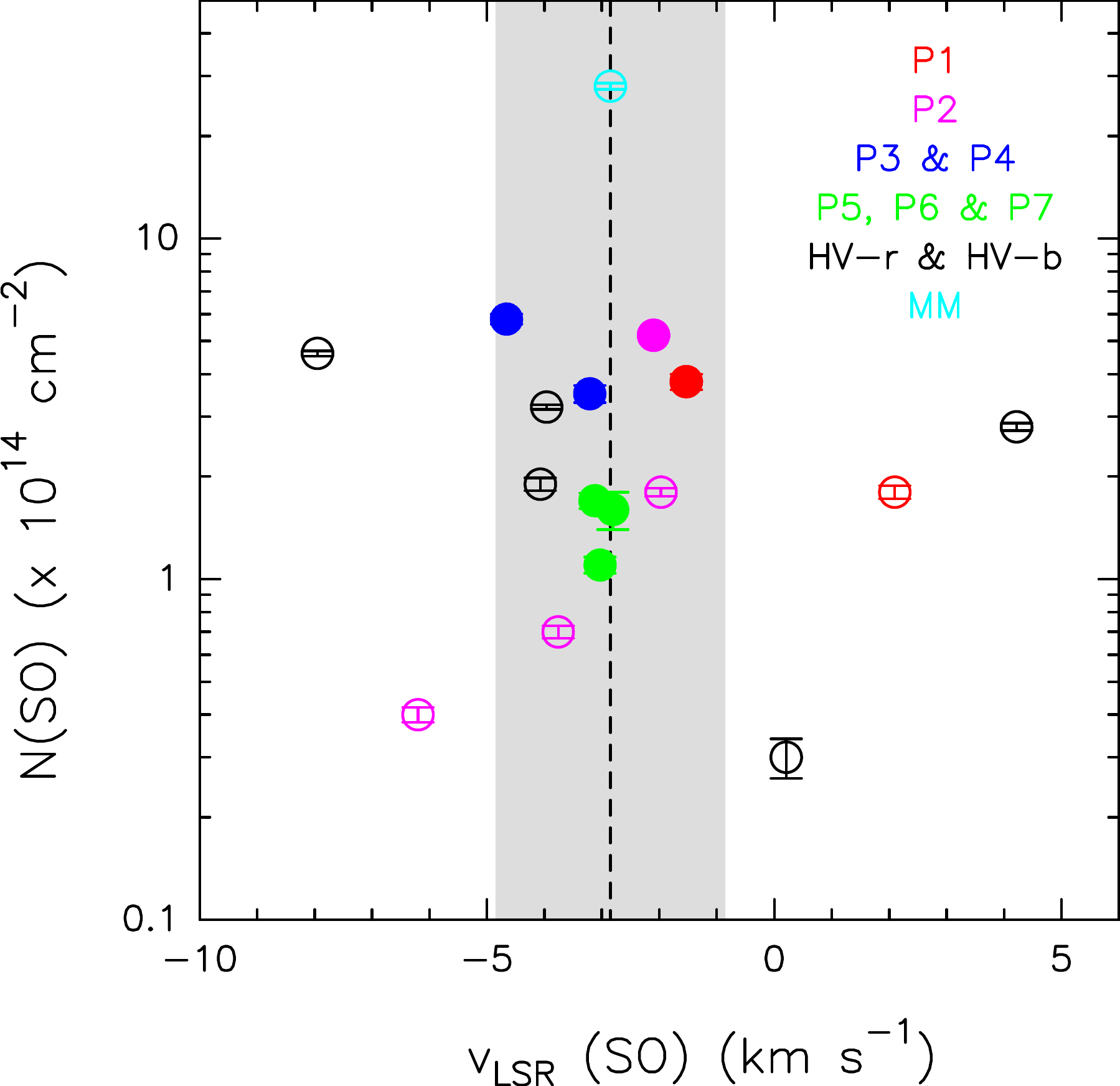}
 \caption{SO column density versus SO velocity towards the different positions in the AFGL5142 field studied in this work. The filled/empty circles indicates detections/non-detections of PN. The different colors denote different group of positions, as labeled in the upper right corner. The uncertainties on the column densities are indicated with error bars. The uncertainties on the velocities are within the size of the symbols. The systemic velocity of the central core, $-$2.85 km s$^{-1}$, is indicated with a vertical dashed line. The gray band indicates a velocity range of $\pm$2 km  s$^{-1}$ with respect to the systemic velocity, which contains all the regions with PN detections.}
    \label{fig_N_vs_vel}
\end{figure}

%\begin{figure*}
%\centering
%\includegraphics[width=18cm,angle=0]{figure-floris.pdf}
%\caption{Non-LTE results for PO and PN from RADEX. {\bf Floris, here we should show only show PN 2$-$1 and PO, in a single row of panels. Coud you redo the plot using the 3 densities and the \tkin\ = 20\,K ?} 
%{\it Upper panels:} Predicted intensities of the PN $J$=2--1, 3--2, and 6--5 lines ({\bf delete 3-2 and 6-5 in the plots) for \tkin\ = 30\,K and gas densities of \pow{1}{5} (left), \pow{3}{5} (middle) and \pow{1}{6} \ccm\ (right). {\it Lower panels:} Predicted intensities of the PO 109\,GHz lines for \tkin\ = 30\,K and gas densities of \pow{1}{5} (left), \pow{3}{5} (middle) and \pow{1}{6} \ccm\ (right).
%}
%\label{fig-floris}
%end{figure*}   

\subsection{ALMA maps of the AFGL5142 star-forming region}
\label{analysis-ALMA}

\subsubsection{Spatial distribution and kinematics of SO: a bipolar outflow cavity}

% velocity of the source: MM in SO is -2.85 km/s
%Hunter 1999: CH3CN (14E-13) and (12E-11) spectra indicate a gas temperature of approximately 65 K in the innermost core.
% starless core revealed with N2H+ emission (\citealt{busquet2011})
% the 3.4 mm source detected by Hunter et al. (1999) actually consists of five millimeter sources (\citealt{zhang2007}). 
% well-collimated SiO jet detected by Hunter et al. (1999)
%Several millimeter continuum sources have been detected towards the center of the field (\citealt{zhang2007}).
% Brief explanation of the region, extending the information of the intro. Important: include the presence of several protostars in the center, based on observations, MM position, continuum peak.
% maser emission reveal a cluster of H2O and CH3OH masers in an area of ∼5′′ (Hunter et al. 1995; Goddi & Moscadelli 2006; Goddi et al. 2007).
%Our 3mm continuum map highlights the central dusty core (Fig. X). {\it Should we do some detailed analysis of the continuum? We cannot add significant information compared with previous works.}
%{\bf Comment: in the next two paragraphs I discuss the results of SO to put the P-bearing emission in context. Indeed, I propose that there is a single bipolar outflow that explain the observations, instead of a cluster of outflows as proposed in previous works. See Fig. \ref{fig-bipolar-outflow}. I am not fully sure that we need this discussion for the purposes of this paper.}

With the aim of interpreting in the following section the maps of PN and PO, we first discuss here the spatial distribution and kinematics of the 2$_3-$1$_2$ transition of SO. Our maps exhibit SO emission from velocities of $-$12 km s $^{-1}$ to +6 km s$^{-1}$ (the systemic velocity of the central  core is $-$2.85 km s$^{-1}$). We show in Fig. \ref{fig-bipolar-outflow} the maps of SO: panel {\it a} includes all velocities, while panel {\it b} includes only the highest velocities with respect the systemic velocity. The SO emission peaks towards the MM position, but also traces extended structures distributed across the field of view. Since SO is considered a good tracer of shocked material (e.g., \citealt{martin-pintado1992,pineau1993,chernin1994,bachiller1997,podio2015}), we interpret the SO emission in terms of shocked (or post-shocked, see below)  material produced by an outflow driven by a central protostar. 

\citet{zhang2007} and \citet{liu2016} interpreted previous observations of CO, HCN and HCO$^+$ as a system of up to three different outflows arising from the central protostar(s). Although this hypothesis is plausible, we propose that the SO emission is tracing the cavities of a single wide-angle bipolar outflow (see overplotted dashed lines in Fig. \ref{fig-bipolar-outflow}). The redshifted and blueshifted high velocity SO emission (Fig. \ref{fig-bipolar-outflow}b) traces very well the launching point of the molecular outflow, coinciding with the central protostar, and the northeast-southwest direction of the bipolar outflow in the plane of the sky  (see zoom-in view in the central inset of Fig. \ref{fig-bipolar-outflow}b). This direction of the bipolar outflow is also in good agreement with the elongated morphology of the Spitzer-IRAC2\footnote{Spitzer-IRAC2 image obtained from the NASA/IPAC Infrered Science Archive: https://irsa.ipac.caltech.edu/Missions/spitzer.html} emission at 4.5 $\mu$m (inset in Fig. \ref{fig-bipolar-outflow}b), which is a good tracer of shocked H$_2$ emission in protostellar outflows (e.g., \citealt{smith&rosen2005,qiu2008}), and with the SiO bipolar outflow detected by \citet{hunter1999}.
The interaction of the lobes of the outflow with the surrounding gas might be responsible for the gas heating observed by \citet{zhang2002,zhang2007} using several inversion transitions of NH$_3$ (see Fig. \ref{fig-bipolar-outflow}c; Zhang, priv. comm.). The detected hot spots towards the northeast and south/southwest are in good agreement with the direction of the bipolar outflow. This suggests that the mechanical energy of the outflow has opened two cavities in the parental core. SO and NH$_3$ mainly trace the interface of these cavities with the local core gas, while the internal regions have already been swept away (Fig. \ref{fig-bipolar-outflow}a and  \ref{fig-bipolar-outflow}c). We have depicted this scenario in a sketch in Fig.  \ref{fig-bipolar-outflow}d. The relatively low velocities of SO with respect to the systemic velocity, around $\pm$10 km s$^{-1}$, and the fact that the gas in the outflow cavities appears redshifted on one side of the cavity and blueshifted on the other side (compare Fig. \ref{fig-bipolar-outflow}a and the sketch in Fig. \ref{fig-bipolar-outflow}d), suggest that the bipolar outflow is close to the plane of the sky.

\subsubsection{Spatial distribution and kinematics of P-bearing species: PN and PO}

The spatial distribution of the PN(2$-$1) emission is shown in the right panel of Fig. \ref{fig-3colors}, where three different velocity ranges  ([-6.5,-4.5], [-4.5,-2.0] and [-2.0, 0] km s$^{-1}$) are shown. For a direct comparison, we show the integrated maps of SO (left panel of Fig. \ref{fig-3colors}) using the same velocity ranges. Unlike SO, PN(2$-$1) emission does not trace gas at high velocities. This is shown clearly in Fig. \ref{fig-PN-SO-HV}, where we compare the spatial distribution of PN with that of the high velocity gas traced by SO. There is a clear anticorrelation, i.e., the regions with high velocity gas traced by SO are devoid of PN emission. 

%We wil discuss the possible implications of this in the
%Indeed, the left panel of Fig. \ref{fig-3colors} shows that 
%has been previously detected in other molecular species such as CO (\citealt{zhang2007}) or HCN and HCO$^+$  (\citealt{liu2016}), and
%Indeed, unlike PN, there is significant SO emission at higher velocities, spanning from -12 km s$^{-1}$ to 6 km s$^{-1}$, as shown in Fig. \ref{fig-bipolar-outflow}a$\&$b. 
The right panel of Fig. \ref{fig-3colors} shows that PN is distributed throughout the field peaking towards several visually-identified spots (P-spots, hereafter) located along the cavities of the bipolar outflow.
% and do not peak towards the MM position where the protostar(s) are nor the starless core.
We have identified seven P-spots in the PN map: P1 (in the northern outflow cavity), and P2 to P7 (in the southern cavity). 

The spectra extracted in a circular region of 3$^{\prime\prime}$ diameter centered on the positions of the P-spots are shown in Fig. \ref{fig-spectra}. 
PO is detected in the P3, P4 and P5/P6/P7 regions, and tentatively in the P2 spot.
In Fig. \ref{fig-PO} we show a zoom-in view for those P-spots for which clear PO emission has been detected. The emission of PN, PO and SO are spatially coincident in the P3 and P4 spots. In the case of the P5/P6/P7 spots, the PO spatial distribution is more similar to that of SO than to that of PN. P-bearing molecules and SO have not been detected towards the position of the starless core (SC).

%{\bf SO is tracing shocks, apparently PO seems to prefer shocked locations compared to PN.}
%{\bf You can also compare with the NH3 temperature map, to see if indeed PO likes warmer regions than PN.}
%{\bf Should we comment more about this? It is interesting that PO seems to be more similar to SO than PN, both spatially and kinematically (see later).}  
%{\it Comment about the spatial coincidence (or not) of PO and PN. In the outer part of the primary beam, the integrated map of PO do not look very well due to poor signal to noise ration}.
%The coordinates of the position of the P-spots are shown in Table \ref{table-sample}. 

We have also extracted spectra towards three additional positions in the region: the central MM position, and the peaks of the redshifted and blueshifted outflow lobes traced by the high velocity SO emission (HV-r and HV-b, respectively, hereafter; see Fig. \ref{fig-PN-SO-HV}). The coordinates of all the positions are indicated in Table \ref{table-positions}. PN is barely detected in the surroundings of the MM position. PO is not detected at any of these positions. None of the P-bearing species is detected toward the HV-r and HV-b. 
The rightmost panel of Fig. \ref{fig-spectra} also shows the spectral profile of the SO 2$_{3}-$1$_2$ transition. It is detected towards all the regions, being strongest at the MM position. The SO spectral profiles show several velocity components in some of the regions. In Section \ref{section-analysis-spectra} we will analyze the spectra towards the different positions to derive the linewidths, velocities and column densities of the different species.

%Fig. \ref{fig-3colors} shows that the PN emission share the same kinematics of SO (same colors in the maps).  

% Comment PN maps in 3 colors
% Identification of PN knots
% Maps of PO in the knots
% Extraction of the spectra in the P-knots

\subsubsection{LTE analysis}
\label{section-analysis-spectra}

%We identified regions (see positions in Table \ref{table-sample}): the P-knots P1-P7, the positions of the central protostar(s) (MM) and 
%%% HIGH VELOCITY POSITIONS
%Besides the positions of the P-spots, the starless core and the MM position, we have also extracted the spectra towards a circular region of 3$^{\prime\prime}$ centered at the peaks of the redshifted/blueshifted high velocity wings of SO. The coordinates of this positions are indicated in Table \ref{table-sample}  and in the left panel of Fig. \ref{fig-PN-noHV}.

We have analysed the spectra of PN, PO and SO extracted towards the different positions using two complementary analyses. In the first case, we have assumed Local Thermodynamic Equilibrium conditions (LTE), as we did in previous studies (\citealt{fontani2016,rivilla2016,rivilla2018,mininni2018}). We have used the Spectral Line Identification and Modeling (SLIM) tool of MADCUBA, which produces synthetic spectra of the molecules using the information from the publicly available spectral catalogs. For the analysis in this work, we have used the entries from the CDMS catalog (\citealt{muller2001,muller2005}). For PN our observations fully resolve for the first time the $^{14}$N hyperfine splitting (see Fig. \ref{fig-spectra})\footnote{The hyperfine structure of PN(2-1) was marginally resolved in single-dish spectra of AFGL 5142 by \citet{fontani2016} (see their fig. 1).}. For this reason we have used the CDMS entry considering the $^{14}$N hyperfine splitting, and the partition function that considers the spin-multiplicities.
%The JPL entry of PN contains the hyperfine transitions, and the partition function includes the spin-multiplicities of $^{14}$N.

To derive the physical conditions, we have used the MADCUBA-AUTOFIT tool that compares the observed spectra with the LTE synthetic spectra of the different species, taking into account all transitions considered. This tool provides the best non-linear least-squared fit using the Levenberg-Marquardt algorithm. The free parameters are: total column density ($N$), excitation temperature ($T_{\rm ex}$), velocity ($\varv$), and full width at half maximum ($FWHM$). MADCUBA-AUTOFIT calculates consistently from these parameters the line opacity of each transition (see \citealt{rivilla2019a}). Since our data contain only a single rotational transition of the three species (PN, PO and SO), the excitation temperature cannot be derived, and hence we assumed fixed values. For the P-bearing species, we used the value found by \citet{mininni2018} in this region using several transitions of PN, namely $T_{\rm ex}$=5 K. We note that the column densities of PN and PO can vary by a factor of 2.5 and 2.1, respectively, assuming different values of $T_{\rm ex}$ in the range of 5$-$50 K (lower column for higher temperatures). For SO, for which an earlier estimate of $T_{\rm ex}$ is not available, we used the kinetic temperature derived from observations of NH$_3$ (\citealt{busquet2011}) of 34 K. The SO column density can vary by a factor of 1.4 assuming a different $T_{\rm ex}$ in the range of 10$-$50 K.  
Then, fixing $T_{\rm ex}$, we run MADCUBA-AUTOFIT leaving $N$, $\varv$, and $FWHM$ as free parameters. Towards some positions (e.g., P1, P2, HV-r and HV-b), we have used several velocity components to reproduce the profile of SO, fixing the value of the velocity as well. 
When the algorithm did not converge, we fixed manually the velocities and/or the FWHM to the values that best reproduced the observed spectra, and reran AUTOFIT. The results of the best fits are plotted in Fig. \ref{fig-spectra} and the derived physical parameters are shown in Table \ref{table-sample}. The errors of the free parameters are derived from the diagonal elements of the covariance matrix, the inverse of the Hessian Matrix, and the final $\chi^2$ of the fit. 

In the following, we summarize the main results of the analysis:

% Comments on the results of the P-spots
$\bullet$  The velocities of PN and PO are different. In the four P-spots for which the two species have been detected and the velocity was left as a free parameter (P3, P4, P5 and P6, all located in the southern blueshifted cavity) the differences are 0.4$\pm$0.1,  0.46$\pm$0.07, 0.24$\pm$0.09 and 0.5$\pm$0.1 km s$^{-1}$, respectively. Fig. \ref{fig_vel_vs_vel} shows that the velocities of PN are lower than those of PO, namely PN is blueshifted with respect to PO. The velocities of PO are more similar to those of SO. 
%{\bf Paola, could we explain this using a chemical argument? Why PO seems to be more kinematically similar to SO than PN?}

%%%%%%%%%%%%%%%%%%%%%%%%%%%%%%%%%%%%%
\begin{table*}
%\scriptsize
\centering

\caption{Results of the non-LTE analysis of PN and PO towards the P-spots where both species have been detected. The column densities have been obtained assuming $T_{\rm kin}$=20 K and three different gas densities $n$ (10$^{4}$, 10$^{5}$ and 10$^{6}$  cm$^{-3}$).}
\tabcolsep 5pt
\begin{tabular}{c c c c c c c c c c c c } 
    \hline
Region   		   & \multicolumn{3}{c}{$N$(PN) (10$^{12}$ cm$^{-2}$)} & & \multicolumn{3}{c}{$N$(PO) (10$^{12}$ cm$^{-2}$)} & & \multicolumn{3}{c}{[PO/PN]$_{\rm non-LTE}$}  \\  \cline{2-4}  \cline{6-8}  \cline{10-12}

%  		   & \multicolumn{3}{c}{$n$ (cm$^{-3}$) } & & \multicolumn{3}{c}{$n$ (cm$^{-3}$)} & & \multicolumn{3}{c}{$n$ (cm$^{-3}$)}  &  \\  \cline{2-4}  \cline{6-8}  \cline{10-12}

 	   & $n$=10$^{4}$  &  $n$=10$^{5}$  &   $n$=10$^{6}$  & &  $n$=10$^{4}$  &  $n$=10$^{5}$  &  $n$=10$^{6}$  & &  $n$=10$^{4}$ &  $n$=10$^{5}$ & $n$=10$^{6}$     \\    \cline{2-4}  \cline{6-8}  \cline{10-12}
 
P2  & 	26.9   & 4.5  & 4.0 & & 39.0 & 8.0  & 9.0 & & 1.4 & 1.8  & 2.2     \\ 
P3  &  14.7   & 2.4  & 2.0   & & 	39 & 8  & 9 & & 2.7 & 3.3  & 4.5     \\ 
P4  &   44  & 5.7  & 4.6   & & 113 & 21  & 21 & & 3.1 & 4.4 & 5.5    \\ 
P5  &  	17.0  & 2.5  & 2.1 & & 82 & 16 & 16 & & 4.8 & 6.4 & 7.6    \\ 
P6  &  16.1 & 2.4 & 2.0 & & 57   &  11 & 12 & & 3.5 & 4.6 & 6.0      \\ 
  
        \hline
	\end{tabular}
\label{table-nonLTE}
%$^{(a)}$ Opacity of the hyperfine transition with highest  $\tau$.
\end{table*}

%%%%%%%%%%%%%%%%%%%%%%%%%%%%%%%%%%%%%

$\bullet$ Most of the P-spots have narrow linewidths in the range 0.9$-$1.7 km s$^{-1}$ (P3, P4, P5, P6 and P7), with the exception of P2, which has broader linewidths of 5$-$6 km s$^{-1}$. The narrow linewidths allowed us to fully resolve for the first time the $^{14}$N hyperfine structure of PN towards the P3 and P4 spots (Fig. \ref{fig-spectra}). 
The quite narrow linewidths of PN and PO towards several P-spots, with values of 1$-$2 km s$^{-1}$, suggest that these molecular species might arise from post-shocked material, where turbulence has been already partially dissipated, rather than directly shocked material. A second alternative would be slow shocks due to the outflow {\it wind}, whose velocity can be significantly reduced compared to the one of the outflow axis, as observed.

%the bulk of the shock, but narrow layers of the gas in the cavity walls.
%{\bf Compare the linewidths of PN, PO and SO}

%The fits towards these positions (assuming $T_{\rm ex}$=5 K) give line opacities of $<$X and $<$0.25 for P3 and P4, respectively. We note that these values are upper limits given that higher values of $T_{\rm ex}$ give lower values of the line opacity.

$\bullet$  The derived line opacities of PN, PO and SO (see Table \ref{table-sample}) are low ($\tau<$0.1) in most of the positions. Only at P4 the transitions are slightly more optically thick, with $\tau\sim$0.2 (for PN and SO), while towards the central MM position the opacity of SO is $\tau=$0.278$\pm$0.009.

%% Comments on the high velocity bullets
%$\bullet$ peaks of the high velocity lobes.

$\bullet$  In the positions where the two P-bearing molecules have been detected, the molecular abundance ratio PO/PN is always larger than 1, with the only exception of the P-spot located closer to the protostar, P2, where the ratio is 0.6$\pm$0.2. In the other P-spots the ratio spans from 1.4 to 2.6 (Table \ref{table-ratios}). This confirms previous findings from IRAM 30m single-dish observations in the W51 and W3(OH) massive star-forming regions (\citealt{rivilla2016}), the protostellar shock L1157 (\citealt{lefloch2016}), and the Galactic Center G+0.693 cloud (\citealt{rivilla2018}), where the PO/PN ratio ranges from 1.8 to 3. 

%% Comments on the results of MM
%$\bullet$  Although PN do not peak towards the central protostar(s), some PN emission is present in the 3$^{\prime\prime}$ region centered at the MM position, as shown in Fig. \ref{fig-ROSINA-spectra}. This emission is likely arising from shocked gas in the surroundings of the protostars.  

$\bullet$ The abundance SO/PN ratio spans from 70 to 207 in the P-spots. These values are orders of magnitude lower than the one at the MM position, which is 3365, and those towards HV-r and HV-b, which are $>$1000 (Table \ref{table-ratios}).  The SO column densities towards these regions are not affected by optical depth effects, since the opacities derived from our analysis are $\leq$0.3 (Table \ref{table-sample}). In any case, we note that even in the case of optically thick conditions, the SO column densities would be lower limits, and thus also the SO/PN value would be a lower limit. Therefore, the SO/PN values found in the P-spots are significantly lower than those of other regions with strong SO emission. 
Comparison between the low value of the SO/PN ratio found in the P-spots and those of the central MM position demonstrates that, while SO traces not only the outflow cavities but also the central core surrounding the protostar (indeed SO peaks towards the MM position), PN emission traces only the cavities. This suggests that P-bearing molecules are excellent tracers of outflow cavity walls.
The SO/PO ratio, shown in the last column of Table \ref{table-ratios},  ranges between 27 and 309 in the P-spots, while it is $>$ 4500 at the MM position. 
%This is due because, while SO traces not only shocked gas but also the central core (indeed SO peak towards the MM position), PN emission highlights only the shocked knots. Therefore, these observations are suggesting that P-bearing molecules are excellent tracers of shocks.

$\bullet$ As previously mentioned, PN and PO are only detected at velocities relatively close to the systemic velocity of the core ($\pm$ 2 km s$^{-1}$), while SO traces also gas at high velocities (Fig. \ref{fig-PN-SO-HV}), as can be seen in the spectra towards P1, P2, HV-r and HV-b (Fig. \ref{fig-spectra}).
%The P-spots are also regions with SO emission, however, some strong regions in SO are not detected in PN or PO, for intance the 
To study this in more detail, we compare in Fig. \ref{fig_N_vs_vel} the column density versus the velocity of SO in the regions considered in this work: the P-spots, MM, HV-r and HV-b positions. The P-spots have SO column densities in the range $\sim$(1$-$10)$\times$10$^{14}$ cm$^{-2}$, and SO velocities ranging from $\sim$ $-$5 to $-$1 km s$^{-1}$. Other regions with comparable SO column densities, but higher velocities with respect to the systemic velocity, do not exhibit emission from P-bearing species.
%However, other regions with similar SO column densities but but high velocities with respect to the systemic velocity do not exhibit P-bearing emission. 
These regions include the HV-r and HV-b positions, and also additional velocity components of the P-spots in which emission of P-bearing species is not detected (Table \ref{table-sample}).  Therefore, the non-detection of P-bearing species in the high-velocity SO spots is not due to a lower column density of SO. 

%{\bf{In summary, P-bearing species are tracing low velocity gas in the cavity walls. This evidence, along with the narrow linewidths, could indicate that P-bearing molecules (and also the low-velocity SO) are tracing post-shocked material, where turbulence has been already partially dissipated, rather than directly shocked material. A second alternative would be slow shocks due to the outflow wind, whose velocity can be significantly lower than that of the outflow axis, as observed}}.
% The regions with <10$^{14}$ cm$^{-2}$.

%{\bf{In the cavity walls, the low velocities of SO with respect to the systemic velocity ($\sim$ $\pm$ 3 km s$^{-1}$)  can be indicative of slow shocks due to the outflow wind, whose velocity can be significantly lower than that of the outflow axis.}}

%Since previous observations of PN and PO (\citealt{mininni2018,rivilla2018}) suggest that these species might be sub-thermally excited, we have also done a non-LTE analysis using the RADEX\footnote{http://home.strw.leidenuniv.nl/~moldata/radex.html} (\citealt{vandertak2007}) code. SO was already available in the code, with collisional rates from \citet{lique2006}. 

%{\color{red} When the algorithm did not converge, we fixed manually the velocities and/or the FWHM to the values that best reproduced the observed spectra, and rerun AUTOFIT. In some cases,  the value of $T_{\rm ex}$ was also fixed (see below). When convergence was not possible, we selected by-eye the solution that best fits the spectra.
%The physical parameters derived are shown in Table \ref{table-parameters}.  

\subsubsection{Non-LTE analysis}

% model setup
Previous works have assumed LTE conditions to estimate column densities of PN \citep{ziurys1987,yamaguchi2011,fontani2016,mininni2018,rivilla2018} and PO \citep{rivilla2016,lefloch2016,rivilla2018}.
However, this assumption may not hold, because the gas densities in the observed regions may not be high enough for the collisional excitation of the lines to compete with radiative decay, and to maintain a Boltzmann distribution of the energy levels.
To test the LTE assumption for the mm-wave lines of PN and PO, we used the non-LTE radiative transfer program RADEX\footnote{\tt https://personal.sron.nl/$\sim$vdtak/radex/index.shtml} \citep{vdtak2007}.
This program solves for the radiative and collisional (de)excitation of the molecular energy levels, and treats optical depth effects with an escape probability formalism.
For PN, we implemented in RADEX the collisional coefficients from \citet{tobola2007}. 
For PO, we used the cross-sections derived by \citet{lique2018} to calculate the collisional rates, which are presented in Appendix \ref{appendix-PO-collision-rates}. 
%We have included the PN and PO files in the LAMDA\footnote{Leiden Atomic and Molecular Database; http://home.strw.leidenuniv.nl/~moldata/} database (\citealt{schoier2005}).
%SO was already available in the code, with collisional rates from \citet{lique2006}. 

%molecular data
%Spectroscopic data for the rotational lines of PN and PO were taken from the CDMS \citep{mueller2005}\footnote{\tt http://www.cdms.de} catalog ({\bf Floris, I realise now that I used JPL because it contains the hyperfine structure of PN, unlike CDMS. Should we use JPL here as well to be consistent?}).
The collisional data for the PN-He system \citep{tobola2007} were scaled by 1.385 to account for \hh\ as the dominant collision partner. 
These data cover the lowest 31 rotational levels of PN, up to 723\,\rcm\, for temperatures between 10 and 300\,K. 
For PO, the recent collision data with He (described in Appendix \ref{appendix-PO-collision-rates}) were scaled by 1.4 to mimic \hh. 
These data cover the lowest 116 rotational hyperfine levels of PO, up to 296\,\rcm\, for temperatures between 10 and 150\,K. 
Both datafiles are available on the LAMDA\footnote{\tt http://home.strw.leidenuniv.nl/$\sim$moldata/} database  (\citealt{schoier2005}).

\begin{figure*}
\centering
\hspace{-19mm}
\includegraphics[width=19.5cm]{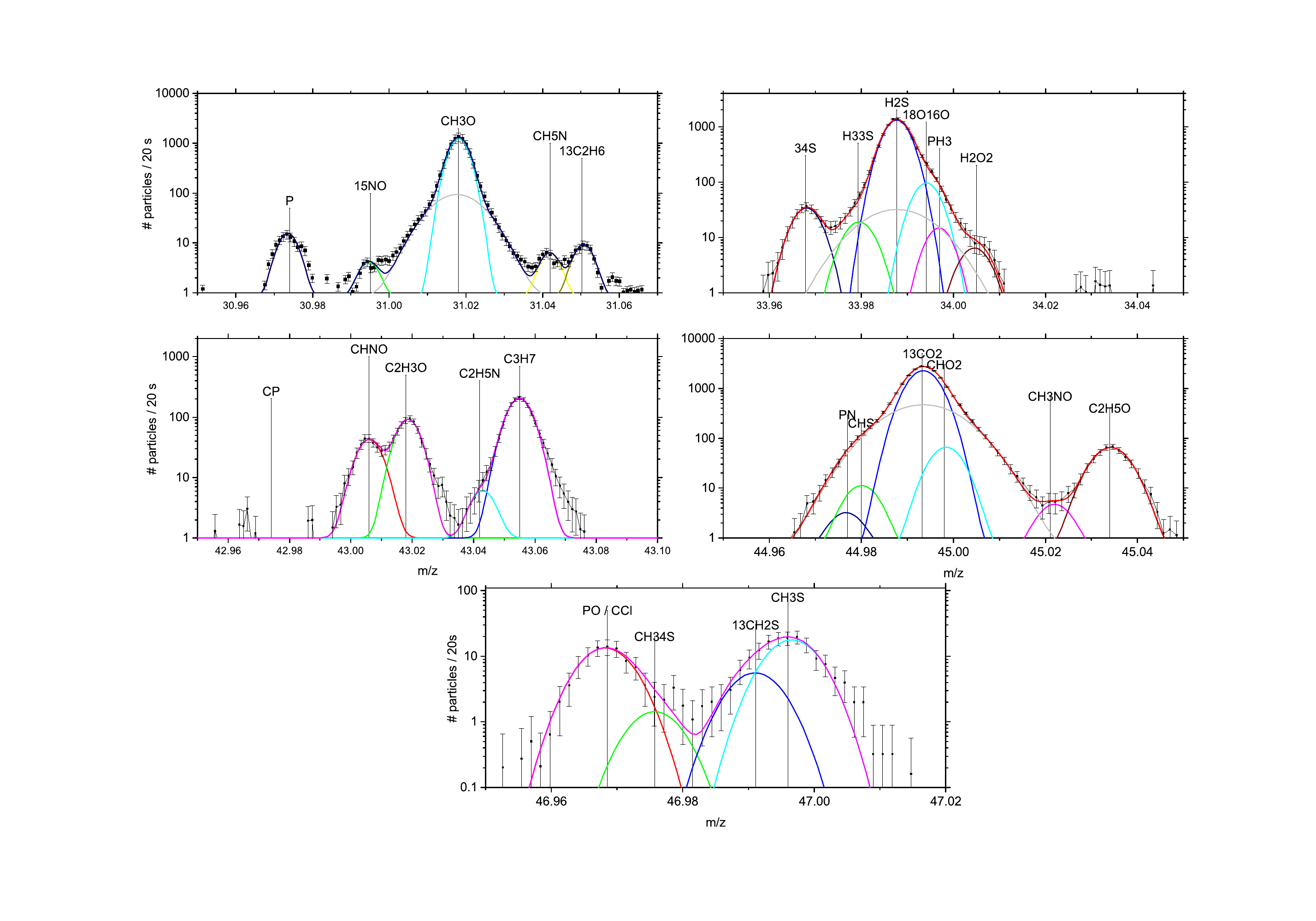}
\vskip-12mm
 \caption{ROSINA DFMS mass spectra for masses 30.9737 Da (P), 46.9681 Da (PO), 33.9967 Da (PH$_3$), 42.9732 Da (CP) and 44.9763 Da (PN). The integration time is 20 s per spectrum. Error bars represent 1$\sigma$ counting statistics. The colored curves correspond to the fit of different species .The P peak was already identified by \citet{altwegg2016}. Regarding the P-bearing molecules, only PO shows a distinct peak at the correct location (see text for the discussion of possible contribution from CCl). For PH$_3$, there is a strong overlap with $^{18}$O$^{16}$O and with abundant H$_2$S. PN has an overlap with CHS, and CP cannot be detected.}
    \label{fig-ROSINA-spectra}
\end{figure*}

%Fig.~\ref{fig-floris} shows the predicted intensities of the PN $J$=2--1 and the PO J=5/2$-$3/2, $\Omega$=1/2, F=3$-$2, l=e transitions., calculated with RADEX as a function of the column density, for \tkin\ = 20\,K and \nhh\ = \pow{1}{4}, \pow{1}{5} and \pow{1}{6}\,\ccm.
%The calculations assume a line width of 1\,\kms\, and a background temperature of 2.73\,K.
%The line width and background temperature are the same as assumed above.

%{\bf We present the observed line intensities in Jy while RADEX give them in temperature units. We should use the same. Floris, could you convert to Jy?}
%Since there is not a direct measure available of the volume density of the P-spots, 
To study the excitation conditions of the P-bearing molecules as a function of the volume density of the gas, we have run RADEX assuming three different values: \nhh\ = \pow{1}{4}, \pow{1}{5} and \pow{1}{6}\,\ccm. The calculations assume a kinetic temperature of \tkin\ = 20\,K and a background temperature of 2.73\,K. For each volume density,  we have then derived the column density that produces the observed line fluxes (in units of K km s$^{-1}$) of the PN $J$=2--1 and the PO J=5/2$-$3/2, $\Omega$=1/2, F=3$-$2, l=e transitions towards the P-spots where both species are clearly detected. The results are shown in Table \ref{table-nonLTE}. 
The derived column densities 
%at  \nhh\ = \pow{1}{6} cm$^{-3}$ are in good agreement with those derived with the LTE analysis (Table \ref{table-sample}), while they 
increase for densities \nhh\ = \pow{1}{5} and especially \nhh\ = \pow{1}{4} cm$^{-3}$ due to strong sub-thermal excitation effects. 
%This means that the abundances of P-bearing species can be 
We also show in Table \ref{table-nonLTE} the PO/PN ratio derived at the different densities. 
%The ratios obtained at \pow{1}{6}\,\ccm \, are consistent with those derived with the LTE analysis (last columns in Table \ref{table-nonLTE}) within the uncertainties. 
Lower gas densities produce lower ratios, but in all cases PO/PN is $>$1, reaching values up to 7.5 in the P5 spot.

\begin{figure*}
%\centering
\hskip-10mm
\includegraphics[width=18.5cm]{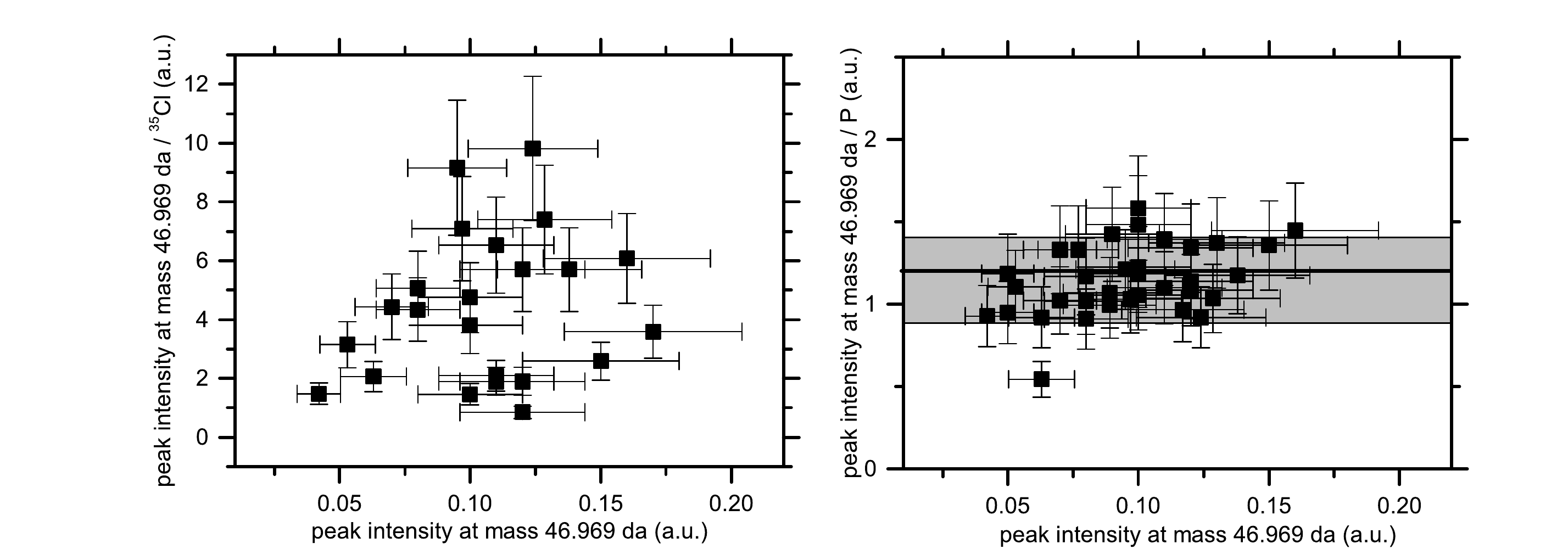}
\caption{Correlation plots between the peak at mass 46.696 Da and $^{35}$Cl (left panel) and P (right panel) as a function of the peak intensity, from ROSINA measurements of the comet 67P/C-G. In the right panel, the black solid line indicates the mean value of the different points, and the grey area denotes the uncertainty of this mean.}
    \label{fig-ROSINA-correlation}
\end{figure*}

\begin{figure}
%\centering
%\includegraphics[width=8cm]{fig-ratio-PO-SO.pdf}
\hskip-10mm
\includegraphics[width=10.5cm]{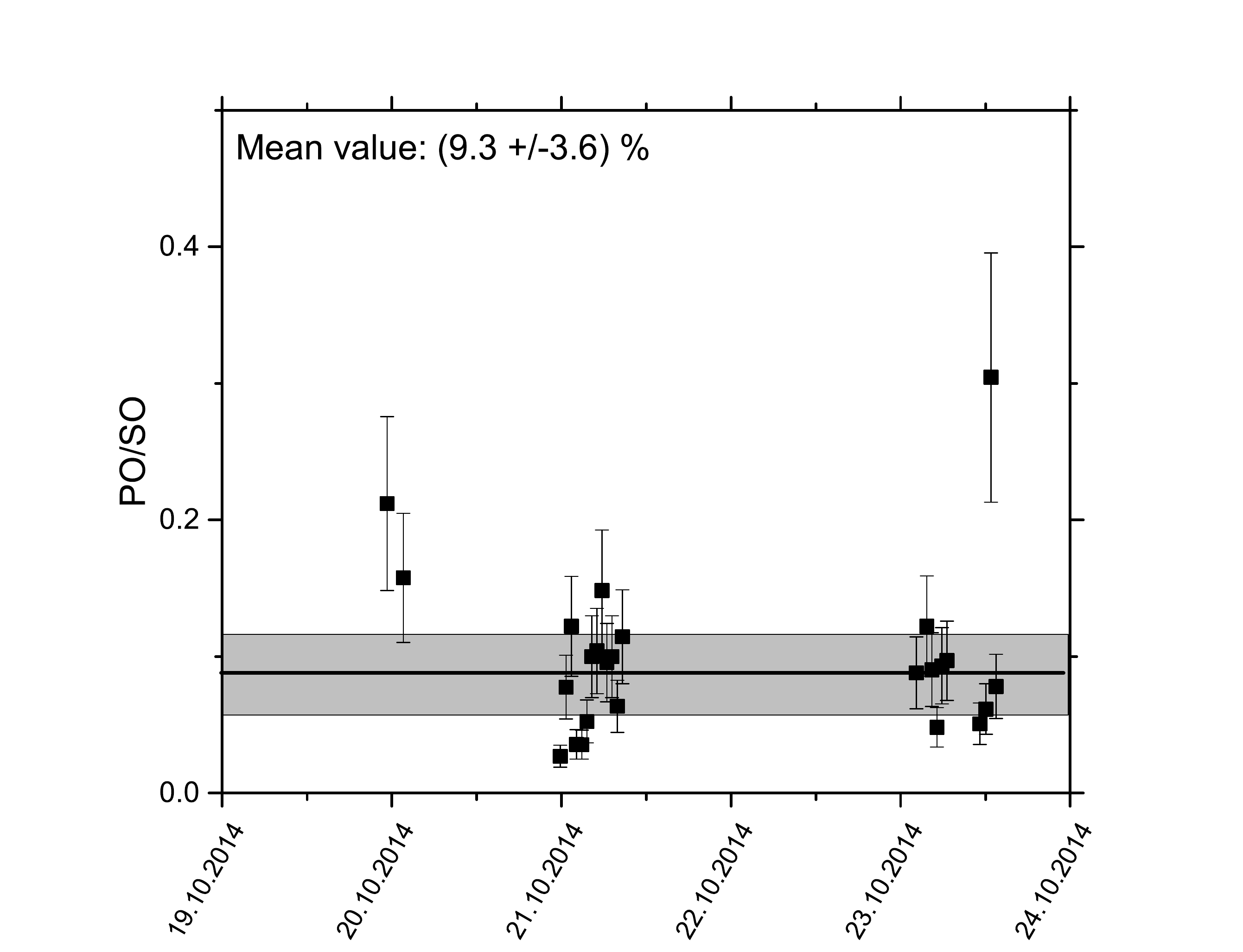}
 \caption{Ratio of PO/SO count rates as a function of time from ROSINA measurements of the comet 67P/C-G. The black solid line indicate the mean value of the different points, and the grey area denotes the uncertainty of this mean.}
    \label{fig-ROSINA-ratio}
\end{figure}

\subsection{ROSINA measurements of the comet 67P/C-G}
\label{analysis-ROSINA}

\citet{altwegg2016} reported the detection of a clear signal on mass 31 Da attributable to P in the comet 67P/C-G using ROSINA data, although it was not possible to detect the parent molecule(s) at that time due to mass line interferences. A re-evaluation of the data has now given more insight into the chemistry of P. The only time when there was a clear detection of atomic P on mass 30.9737 Da was when {\it Rosetta} orbited in the terminator plane at 10 km from the nucleus centre from October 14 to October 29 2014. Before and after this time, {\it Rosetta} was too far from the comet and/or the production rate was too low. A full orbit around the comet took $\sim$2.5 Earth days, which means 5 full orbits for the whole period. ROSINA obtained $\sim$1 spectrum/hour for $\sim$20h/day for the full mass range. So the number of spectra is on the order of 200 spectra per mass.  In about 1/3 of them P could be detected in a statistically significant amount due to the large diurnal and seasonal density variations. To look for possible parents of P, beyond mass 31 Da, we have also analysed masses 34 Da for PH$_3$, 43 Da for CP, 45 Da for PN, 47 Da for PO. Sample spectra for 31, 34, 43, 45, and 47 Da are given in Fig. \ref{fig-ROSINA-spectra}. 

The P peak on mass 31 Da is easy to detect, as it is separated from other contributions on mass 31 Da. PH$_3$ is completely hidden beneath the peaks of $^{18}$O$^{16}$O and abundant H$_2$S. No signal could be detected at the mass of CP, which would be well separated from interfering peaks. PN is completely hidden by the peaks of CHS and $^{13}$CO$_2$. 

At the mass of PO (46.696 Da), there is a clear peak. However, PO and CCl strongly overlap because their masses are identical up to the fourth digit. In order to see if the peak at 46.969 Da is due to CCl or PO or both, we checked the correlation of the ratio of this peak with the P peak on mass 31 Da and with the $^{35}$Cl peak on mass 35 Da. The result is shown in Fig. \ref{fig-ROSINA-correlation}. The ratio m$_{46.696}$ / $^{35}$Cl or m$_{46.696}$ / P, respectively, should be nearly constant whether the peak on mass 46.696 Da is due to CCl or PO, respectively. This is the case for P, but not for $^{35}$Cl. A possible parent of CCl could be CH$_3$Cl. According to NIST (\citealt{stein2016}), the fragment CCl from CH$_3$Cl is about 7$\%$ of the parent molecule. This would mean that the signal for CH$_3$Cl should be well above 100 particles / 20 s. No signal at all was detected on mass 50 Da of CH$_3$Cl, although later in the mission there is clear evidence for CH$_3$Cl (\citealt{fayolle2017}).  CH$_3$Cl seems to be correlated to dust which was not abundant in October 14.
%{\bf Question for Kathrin and Maria: does this mean that CH$_3$Cl is located in another region of the comet?}
% KA: No, it probably means that sublimation of CH3Cl needs higher temperatures (heliocentric distances) than for PO. 
We therefore exclude the presence of CCl and attribute all of the peak at mass 46.696 Da to PO.

In order to derive a relative abundance of PO, we have to make a few assumptions:
i) we assume that the ionization cross section of PO is the same as for NO as there exist no data for PO and it cannot easily be calibrated in the lab due to its unstable and possibly poisonous nature;
ii) we assume the same instrument sensitivity for SO and PO as they are very similar in mass, which cancels out the mass dependent sensitivity of the instrument.
%we assume that PO correlates with SO. The reason to do so is that SO and PO are close in mass, so that the instrument sensitivity does not play a role.
%{\bf Question for Kathrin and Maria: I do not fully understand this. Could we explain it better?} 

%KA: this we don't assume, but from the data we have they are more or less proportional, which means they are embedded similarly in the ice and have similar sublimation temperatures which depend mostly on the matrix they are embedded in.

% KA: the measured abundance in a mass spec is given by the ionization cross section, the fragmentaztion pattern (e.g. PH3 gives you PH3+, PH2+, PH+ and P+; PO gives you PO+ and P+ and O+) and the instrument sensitivity. The latter is mass dependent. But PO and SO are very close in mass, so the latter does not play a role.

Fig. \ref{fig-ROSINA-ratio} shows the correlation between PO and SO. In order to get the quantity of SO we also have to take into account SO$_2$ and then subtract the fragments from SO$_2$ from the signal on SO (for details see \citealt{calmonte2016}). SO and PO correlate well. These measurements were done at 3 au inbound, when there was a large heterogeneity in the coma between CO$_2$, CO and H$_2$O (\citealt{hassig2015}), depending on the sub-spacecraft latitude. SO behaves similar to CO as does PO. A correlation with, e.g., water or methanol, yields huge spreads. The good correlation between SO and PO suggests that they are embedded in a similar way in the ice and have similar sublimation temperatures, which depend mostly on the matrix in which they are embedded.

For other P-bearing molecules, we can only get upper limits. We assume that the ionization cross sections of all the species are similar.  The signal on mass 45 Da gives an abundance ratio of PO/PN$>$10.  An abundance ratio of PO/PH$_3>$3.3 would still be compatible with our mass spectra and a Solar $^{18}$O / $^{16}$O ratio. There is no signal for CP, which means a ratio of PO/CP $>$33. 

The signal at mass 31 Da is a daughter product of PO and possibly PH$_3$. Taking the fragmentation pattern of NH$_3$ as a proxy for PH$_3$, only very little P is produced. That means that most of the peak at mass 31 Da has to be attributed to PO which adds to $\sim$70$\%$. To reach this conclusion, we have taken the mass dependent sensitivity of DFMS into account.  This yields, finally, an abundance of PO relative to SO of 16$\%$ with a probable uncertainty of a factor of 2 due to the assumptions made. This implies a SO/PO ratio of $\sim$6.

While PO was not detected later in the mission, SO is present throughout, due to its higher abundance.  \citet{calmonte2016} have deduced a nucleus bulk abundance for SO/H$_2$O of 7$\times$10$^{-4}$ for the period May/June 2015 just before perihelion, but outside of the dust outbursts of the near perihelion phase (\citealt{vincent2016}). If we assume that PO and SO are indeed well correlated, this would yield a relative abundance for PO/H$_2$O of 1.1$\times$10$^{-4}$ for the nucleus bulk within a factor of 2. 

The P/O ratio found in the comet is 0.5$-$2.7$\times$10$^{-4}$, close to the Solar value of 5.8$\times$10$^{-4}$, indicating that P is only slightly depleted (see also \citealt{rubin2019b}).

 \begin{figure*}
\centering
%  SCRIPT: Speedsheet powerpoint "AFGL5142-photochemistry"
\includegraphics[width=12cm, angle=-90]{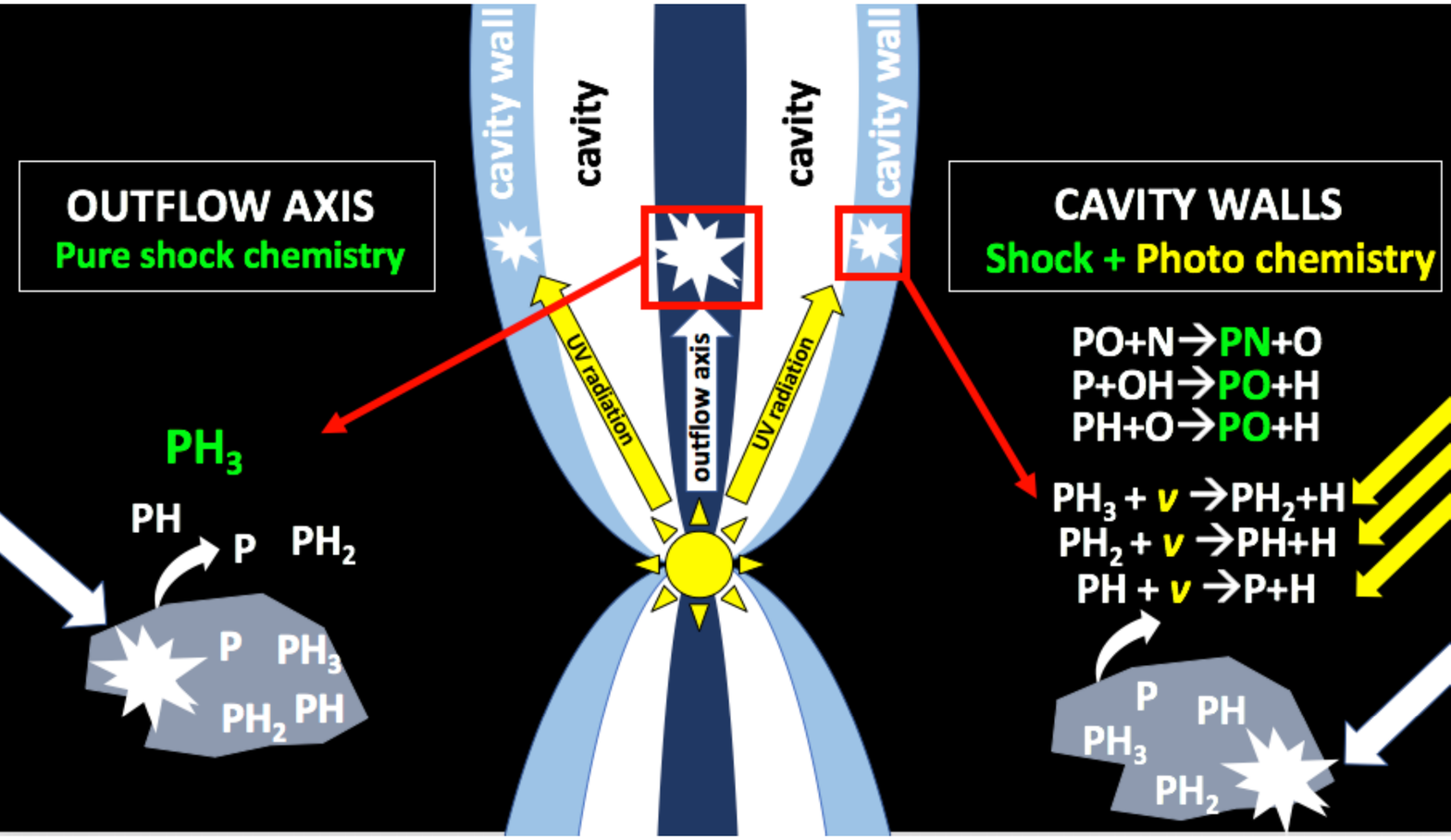}
\vskip-10mm
 \caption{Scenario proposed for the formation of P-bearing species in the cavity walls excavated by a molecular outflow. Along the high-velocity outflow axis, the chemistry is mainly dominated by shocks, and PH$_3$ is expected to be the most abundant P-bearing species. In the UV illuminated cavity walls, photochemistry is producing PO and PN.}
    \label{fig-photochemistry}
\end{figure*}
   
\section{Discussion}
\label{discussion}

\subsection{Formation of P-bearing molecules in star-forming regions: the role of shocks and photochemistry}
\label{formation}

\subsubsection{Hints from the spatial distribution of the emission}

We have presented the first high-angular resolution maps of P-bearing species (PO and PN) in a star-forming region, which allow us to shed light on the poorly constrained chemistry of interstellar P. The non-detection of strong emission of P-bearing species towards the positions of the central hot core (MM position) and the position of the starless core (SC) seems to challenge the formation theories based on hot gas-phase chemistry triggered by the warm-up of protostars (\citealt{charnley1994}), and on cold gas-phase chemistry during the prestellar collapse phase of the parental core (\citealt{rivilla2016}). 
%{\bf Should we mention here that we missed a lot of flux with ALMA and that, as a consequence, we may missed emission from a cold and large envelope surrounding the protostar(s) and/or the starless core?
%These ALMA observations are probing spatial scales up to 15$\arcsec$.}

% temperature (T ≃ 70 K) and high density (n ≃ 106 cm−3) of MM (\citealt{busque2011})
%{\bf Rephrase this paragrapgh}

Previous theoretical models (\citealt{aota2012,jimenez-serra2018}) propose that shocks are the most important agent for the chemistry of P-bearing species. According to this scenario, the ice mantles of the grains are sputtered due to the mechanical energy of the shocks, releasing P into the gas phase. 
The shock scenario has been supported by the detection of PN and PO in a protostelar shock (\citealt{lefloch2016}) and by the observations of \citet{rivilla2018} in a survey of molecular clouds in the center of our Galaxy. In this study, we found that PN and PO are only detected in the regions whose chemistry is dominated by large-scale shocks, and the abundances of PN correlate with those of the typical shock tracer SiO (e.g., \citealt{martin-pintado1992}). This observational correlation between PN and SiO has also been confirmed in a survey of high-mass star-forming regions observed with single-dish telescopes (\citealt{mininni2018}, Fontani et al., submitted), suggesting a link between shocks and the presence of P-bearing species.

However, our high-resolution ALMA observations towards AFGL 5142 have provide us with essential new information that 
 challenges the idea that shocks are the {\em main} driver of P-chemistry. PN and PO have been found only in the cavity walls (Fig. \ref{fig-3colors}), and not along the central high-velocity parts of the bipolar outflow (e.g., HV-r and HV-b positions), where SO is also prominent (Figs. \ref{fig-PN-SO-HV} and \ref{fig_N_vs_vel}). This indicates that the mere presence of shocked material does not guarantee the detection of PO and PN. This may suggest that these molecules are not directly sputtered from grains, but are likely formed later through gas-phase chemistry.

%the merely presence of shocks indicates that
%is not able to explain the observations, 

%and that another agent that is key in the chemistry of P-bearing species
\citet{jimenez-serra2018} studied the chemistry of P-bearing species theoretically, considering not only the presence of protostellar shocks, but also other physical mechanisms such as protostellar heating, UV and cosmic-ray irradiation. 
%In the following we interpret our ALMA results comparing with the predictions of these chemical models.
Based on our ALMA maps of AFGL 5142 and the predictions of these chemical models, we propose that, besides shocks, photochemistry is also key for regulating the abundance of P-bearing species. We depict this scenario in the scheme presented in Fig. \ref{fig-photochemistry}. Along the outflow axis, traced by the high-velocity SO emission, the chemistry is mainly dominated by shocks. In this region, atomic P on the surface of dust grains is efficiently hydrogenated forming species such as PH, PH$_2$ and mainly PH$_3$, which afterwards desorb thanks to shock-induced grain sputtering. This region of high-velocity gas is thought to be younger than the material of the cavity walls, and further gas-phase chemistry is thus not expected because it would need longer timescales. In this scenario, PH$_3$ should be the dominant P-bearing species in the freshly desorbed material, according to the shock model of \citet{jimenez-serra2018}. Since the extinction is higher along the outflow axis than in the direction of the cavity walls, the UV photons are not efficient converting PH$_3$  to PO and PN through gas-phase photochemistry.

%%%%%%%%%%%%%%%%%%%%%%%%%%%%%%%
%PH3, should be the dominant P-bearing species in the freshly desorbed material in the outow axis, because it can be1 formed very eciently on the surface of grains through hydrogenation of atomic P
%%%%%%%%%%%%%%%%%%%%%%%%%%%%%%
In contrast, the situation in the cavity walls is different. Since they are more exposed to the UV photons arising from the central protostar, photochemistry can govern the chemistry. As indicated in Fig. \ref{fig-photochemistry}, the hydrogenated P-bearing species desorbed by shocks (e.g. PH$_3$) can rapidly (on short timescales of 10$^{4-5}$ yr) be converted to PO and PN, according to the gas-phase photochemistry models of \citet{jimenez-serra2018}. The formation of PO and PN is enhanced in the cavity walls, where the volume densities become large enough for formation reactions to proceed rapidly. Thus, the scenario depicted in Fig. \ref{fig-photochemistry}, with a combination of shocks and photochemistry, may explain why PO and PN are only observed in the cavity walls and not along the outflow axis.  A  positive detection of PH$_3$ in the shocked gas along the outflow axis would support this hypothesis. 

It should be noted that photochemistry might be not only productive (converting PH$_{3}$ into PO and PN), but also destructive. Assuming a UV radiation field of 10 $G_{\rm0}$, and using the photodestruction rates proposed by \citet{jimenez-serra2018}, 3$\times$10$^{-10}$ s$^{-1}$ and 5$\times$10$^{-12}$ s$^{-1}$ for PO and PN, respectively, the unshielded timescales for visual extinctions $A_{\rm V}>$5 mag\footnote{The UV radiation field is attenuated due to the visual extinction A$_{\rm V}$ following the exponential decay e$^{-\gamma A_{\rm V}}$. We have used $\gamma$ of 2 and 3 for PO and PN, respectively, as proposed by \citet{jimenez-serra2018}.} are $>$2$\times$10$^5$yr and $>$4$\times$10$^{10}$yr, respectively. So, in principle, P-bearing molecules (including PO that is more easily photodissociated) can survive for relatively long times if they are partially shielded, in agreement with our observations.

\subsubsection{Hints from molecular abundance ratios}

The observed PO/PN molecular abundance ratios are a key additional constraint that supports this scenario. We have found that PO is more abundant than PN   towards most of the P-spots, with the only exception of P2. Previous detections in star-forming regions and the Galactic Center cloud G+0.693 (\citealt{rivilla2016,lefloch2016,rivilla2018,bergner2019}) found also that the PO/PN ratio is $\geq$ 1. 
However, these values $>$ 1 are never produced by the pure shock models of \citet{jimenez-serra2018} (see their Fig. 11). To reproduce the observed PO/PN ratios, these authors offer several alternatives. 
%The first is the inclusion of a new chemical reaction, P+OH $\rightarrow$ PO+H, not yet included in any chemical database. The observational correlation (spatial and kinematical) between PO and SO (Section \ref{analysis}) in the P-knots supports this route, since SO is formed in a similar way in gas-phase: S + OH $\rightarrow$ SO + H. While this reaction could certainly increase the production of PO, it would not explain by itself (without invoking photochemistry) why PO is only formed in the cavity walls and not in the outflow axis. In any case, theoretical calculations and laboratory experiments are needed to confirm the viability of this proposed chemical route.
%have proposed that ratios PO/PN$>$1 in shocks can be obtained with new formation routes of PO that are not currently considered in the chemical models. In particular, they proposed the reaction P + OH $\rightarrow$ PO + H. 
%Including this new reaction, PO can be more abundant than PN in shocks even for standard cosmic-ray ionization rates. However, theoretical calculations and laboratory experiments are needed to confirm the viability of this proposed chemical route.
%We have found that PO is more abundant than PN  (by a factor 2$-$9) towards all the P-spots. 
%In AFGL 5142 we have seen that the emision from P-bearing species arises from shocked gas associated with the bipolar molecular outflow. 
The first one is hot chemistry ($\sim$100 K) produced by protostellar heating. However, our ALMA observations did not detect P-bearing emission in the central hot core-like region surrounding the central protostar, but in the colder cavity walls, ruling out this possibility.

A second possibility proposed by \citet{jimenez-serra2018} is a high cosmic-ray ionization rate. Enhanced values of the cosmic-ray ionization rate with respect the typical interstellar values ($\zeta$=10$^{-17}-$10$^{-16}$ s$^{-1}$, \citealt{indriolo2015}), have been claimed in molecular outflows powered by low- and intermediate-mass stars (\citealt{ceccarelli2014,podio2014}), with values of $\zeta$ spanning from $\sim$10$^{-12}$ to 10$^{-16}$ s$^{-1}$, and in protocluster star-forming regions like OMC$-$2 FIR4 ($\zeta\sim$10$^{-14}$; \citealt{fontani2017}). One may expect a similar effect in a massive star-forming region such as AFGL 5142. These high values might be reached due to the acceleration of both cosmic-ray protons and electrons through diffusive shock acceleration (\citealt{padovani2016}). However, a high cosmic-ray ionisation rate is unlikely to be the reason of the PO/PN ratios observed. First, because the models by \citet{jimenez-serra2018} that predict PO/PN$\sim$10 does not only need  $\zeta$=1.3$\times$10$^{-13}$ s$^{-1}$, but also high temperatures $\sim$100 K, which are not present in the AFGL 5142 outflow cavities. Second, while is true that the combination of high $\zeta$ and high temperature produce high PO/PN ratios, the absolute molecular abundances are very low, $\leq$10$^{-12}$ (see Fig. 6 of  \citealt{jimenez-serra2018}), which are below the detection limits. And third, the mechanism proposed by \citet{padovani2016} to increase the cosmic-ray ionisation rate is expected to be mainly relevant along the outflow axis rather than in the cavity walls. 

%Considering $\zeta$=1.3$\times$10$^{-13}$ s$^{-1}$, PO can be a factor of $\sim$10 more abundant than PN. However, 
%However, since this mechanism is expected to be relevant along the outflow axis rather than in the cavity walls, it does not seem the most likely explanation for the PO/PN ratios found in the P-spots.  
%Therefore, the ratios PO/PN$>$1 found towards the P-spots might be a direct consequence of the enhancement of the cosmic-ray ionization rate.

\begin{figure}
\centering
%  SCRIPT: @ratio_vs_distance.greg
\includegraphics[width=7.5cm]{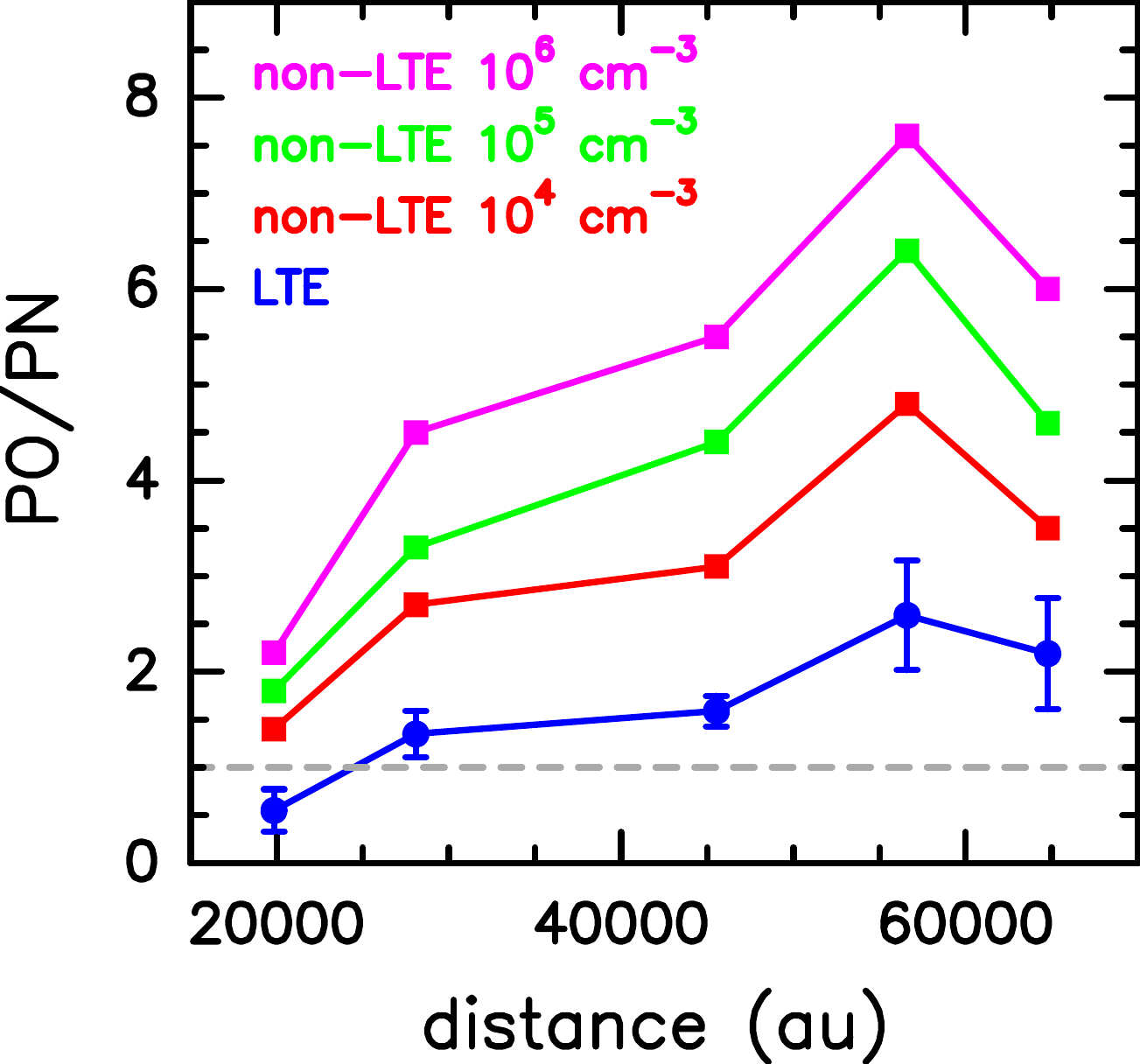}
 \caption{PO/PN molecular ratio, normalized to the ratio found in the P2 spot, as a function of the distance to the central MM position. We have included all the P-spots of the southern cavity, with the exception of the P7 spot, because it is located very close to the primary beam of the observations, where poorer sensitivity may affect the accuracy of the measured PO column density. The different colors indicate the molecular ratios obtained with the LTE approximation (blue dots) and the non-LTE analysis considering different densities: 10$^{4}$ cm$^{-3}$ (red squares), 10$^{5}$ cm$^{-3}$ (green squares), and 10$^{6}$ cm$^{-3}$ (magenta squares).}
    \label{fig_ratio_vs_distance}
\end{figure}

The last alternative to produce high PO/PN ratios is photochemistry. As shown in Fig. 11 of \citet{jimenez-serra2018}, the models based on photochemistry with intermediate extinctions (A$_{\rm v}\sim$7 mag) produce PO/PN$>$1, as observed in the cavity walls of AFGL 5142. This is further support for the photochemistry scenario we are proposing. Moreover, we have seen before that PO is easier to photodissociate because it is thought to have a larger photodestruction rate than PN by almost 2 orders of magnitude (\citealt{jimenez-serra2018}). 
Therefore, PO is expected to be more efficiently destroyed close to the star, where the UV radiation is more intense. Moreover, even if PN is photodissociated, it may be able to reform again (unlike PO) thanks to the large amounts of atomic P available in the gas phase, which react via P + CN $\rightarrow$  PN + C (\citealt{jimenez-serra2018}).
Consequently, the PO/PN ratio is expected to increase with decreasing UV flux, i.e., with increasing distance from the protostar. This theoretical prediction can be checked using our ALMA data. In Fig. \ref{fig_ratio_vs_distance} we show that the PO/PN ratio in the different P-spots of the southern cavity walls of AFGL 5142 (normalized to the ratio found in the P-spot P2), obtained by the LTE and non-LTE analyses. In all cases, the PO/PN ratio clearly increases with increasing distance from the central protostar, supporting the role of photochemistry to explain the observed relative abundances of PN and PO.
We note however that a proper physical modeling of the source coupled with a detailed treatment of photochemistry should be done in the future to fully confirm this scenario proposed here qualitatively.

\subsubsection{Scenario proposed for the formation of PO and PN}

In summary, the scenario that emerges from our ALMA observations of AFGL 5142 explains the formation of P-bearing species in three main phases:

\begin{itemize}

\item Rapid hydrogenation of atomic P on the surface of dust grains, which form efficiently PH$_3$ during the early cold phase.

\item Grain disruption induced by protostellar outflows that desorb PH$_3$ into the gas phase. 

\item Destruction of PH$_3$ and formation of PO and PN through gas-phase photochemistry in the outflow cavity walls, which are exposed to significant UV radiation from the central protostar.

\end{itemize}

Therefore, we propose that PO and PN are formed after the protostellar birth, since they are a consequence of events produced by the protostellar activity: shocks and UV radiation. This scenario may be favoured in regions with high H/H$_2$ ratio, such as those dominated by photons,  which are typical of high-mass star-forming regions like AFGL 5142.

According to the scenario summarized in Fig. \ref{fig-photochemistry}, PO is formed via  PH + O, and also via P + OH. This latter formation route, not yet included in any chemical database, has been recently proposed by \citet{jimenez-serra2018}. Although theoretical calculations and laboratory experiments are needed to confirm the viability of this reaction, we have found observational evidence that supports it. Our ALMA data shows that while PN and PO are systematically shifted in velocity by $\sim$0.5 km s$^{-1}$  (being PN more blueshifted, Fig. \ref{fig_vel_vs_vel}), PO and SO share the same velocity in the different regions studied. Moreover, Fig. \ref{fig-PO}c shows that PO follows the morphology of SO better than PN. Therefore, the observational correlations (spatial and kinematical) between PO and SO in the P-spots support the proposed formation route P + OH $\rightarrow$ PO + H, since SO is formed in a similar way in the gas phase: S + OH $\rightarrow$ SO + H (\citealt{kaufman1996,laas2019}). 
%The OH radical may be more abundant in regions where photons can phodissociate H$_2$O.

The different velocities of PO and PN (Fig. \ref{fig_vel_vs_vel}) suggest that both species trace different layers of the outflow cavity walls, which may have slightly different physical conditions. Observations have shown than oxygen- and nitrogen-bearing species behave very different in relatively cold regions of the ISM, in particular they differ on how they deplete onto dust grains (e.g. \citealt{caselli2017}). At low temperatures (below 20 K) O-bearing species deplete significantly, while N-bearing species do not show evidence of depletion or freeze-out (e.g. \citealt{hily-blant2010,spezzano2017}). Therefore, in the gas PN would survive better than PO in the coldest layers of the outflow cavity walls, those located farther away with respect to the protostar. In constrast, the closest layers of the cavity wall, more exposed to the protostellar heating and then with slightly higher temperatures, would favor the presence of PO (and SO). This could explain the observed differences between the velocities of the O-bearing (PO and SO) and the N-bearing species (PN). We have seen that PN is more blueshifted than PO and SO in the P3, P4, P5 and P6 spots, which are located in the sourthern and blushifted cavity of the outflow. In an accelerating outflow, the higher velocity of PN with respect to the central position may indicate that PN is arising from a deeper (and thus colder) layer in the cavity walls, where O-bearing species are more depleted.

The angular resolution of our ALMA maps ($\sim$2$\arcsec$) is likely not high enough to resolve spatially these different regions in the cavity walls. Fig. \ref{fig-PO} does not show clear differences in the morphology of PO and PN towards the P3 and P4 spots. In the case of the P5/P6/P7 regions, as already mentioned, there is a difference in the spatial distribution of PO (that follows SO) and PN. The latter appears shifted towards the SE, i.e., farther from the protostar. As a consequence, this part of the cavity may be indeed colder than that traced by PO, in good agreement with the selective depletion scenario described before. However, the current data only allows to draw some speculative conclusions, and higher angular resolution observations are needed to test this hypothesis.

\subsection{The chemical thread of Phosphorus}

\subsubsection{From the ISM of star-forming regions to Solar-system bodies}

We have described in the previous section how PO and PN can be synthesized in the ISM. We have seen that a massive star-forming region such as AFGL 5142 is a environment favorable for the formation of these P-bearing species, since it provides the two main physical agents required: protostellar shocks to sputter P from the dust grains and UV photons to trigger the photochemistry that forms PO and PN. The question now is to understand how this P reservoir that is available in the gas phase of a massive star-forming region can be incorporated to a planetary system similar to our Solar System. In this sense, the comparison between the P-content of AFGL 5142 and that of the comet 67P/C-G can give us important clues. 

% ORIGIN OF THE SOLAR SYTEM

It is well known that most stars are not born in isolation, but within clusters (e.g., \citealt{carpenter2000}, \citealt{lada&lada2003}) that also include massive stars (e.g., \citealt{rivilla2013a,rivilla2013b,rivilla2014}). There is multiple and independent evidence that indicate that our Sun was not an exception.
%  (see detailed reviews by \citealt{adams2010} and \citealt{pfalzner2015}).
%and hence that it was formed within an environment with the presence of massive stars. 
Based on the observed extreme orbital eccentricity of the minor planet Sedna and the measurements of short-lived radioactive species (e.g. $^{26}$Al) inferred from meteorites, it has been proposed that our Sun was born in a cluster with 10$^{3}-$10$^{4}$ stellar members (\citealt{adams2010},  \citealt{pfalzner2015}), including massive stars. 
As a consequence, the chemical content of the ISM of the natal (massive) star-forming region might be incorporated into the ProtoSolar Nebula (PSN) and into the Solar System bodies, including comets and our own planet.

The abundances of P-bearing molecules with respect to H$_2$ found in the AFGL 5142 star-forming region are difficult to estimate, since the abundance of H$_2$ in the P spots is unknown. Using the molecular abundances of P-bearing species reported in previous works of 10$^{-9}-$10$^{-10}$ (\citealt{fontani2016,rivilla2016,lefloch2016,mininni2018,bergner2019}), we infer that the P in these molecules account for the 0.03$\%$ - 0.3$\%$ of the cosmic P. This is certainly a small fraction of the total P content in the ISM, although non negligible. This P in the form of simple volatiles, if transferred during the star formation process to Solar System bodies, could be enough to provide a reservoir of P accessible for prebiotic chemistry, creating an environment favorable to foster Life (see further discussion in next section).

Thanks to the {\it{Rosetta}} mission, there is now multiple evidence indicating that the chemical composition of  the comet 67P/C-G was set before the birth of the Sun, and has survived almost unaltered since then. Measurements of N$_2$ (\citealt{rubin2015}), deuterated species (\citealt{altwegg2017_deuterium}), S$_2$ (\citealt{calmonte2016}), Xe (\citealt{marty2017}), and O$_2$ (\citealt{bieler2015}), among others, have shown that the ice composition of the comet is pre-Solar. 
% and thus that ices were assembled.
%Detection of N$_2$ indicates a low formation temperature below 30 K (\citealt{rubin2015}), high deuteration levels (\citealt{altwegg2017_deuterium}) can be explained  if much of the ice survived as ice from prestellar stage to the comet (e.g. \citealt{furuya2016}). 
%The detection fo the very volatile S$_2$ (\citealt{calmonte2016}), which has a very short photo lifetime of $<$ 30 s at 1 au, implies that it was locked on the ice before comet formation. 

Similarly, our analysis of PO in 67P/C-G suggests that it was already in the ice in the prestellar phase. {\em In-situ} photochemistry can be ruled out as the origin of the observed PO, since it was detected at $\sim$10 km from the nucleus at 3 au. This means that the molecules are exposed to a (weak) Solar illumination during only $\sim$10s, so photochemistry cannot take place. The ROSINA data have been examined for larger, more complex P-bearing molecules (such as PO$_2$, PO$_3$, H$_x$PO$_y$), yielding negative results. Moreover, PO is suspected to be strongly correlated with CO$_2$, since it shows higher abundances over the southern (winter) hemisphere, similar to CO$_2$ and contrary to H$_2$O, which was more abundant in the north at that time. This suggests that PO is mixed with CO$_2$ ice. Hence, there is no evidence to suggest that PO stems solely from the surface of the comet, nor it being a fragment of some other P-rich compound.  Therefore, the volatile PO measured by ROSINA arises from ices that were assembled prior to the formation of the comet.  This means that the comet may have directly inherited the chemical composition of its parental PSN, which was originated in a clustered environment with neighboring massive protostars.

%After its formation, it freezed out again into ices, and these ices were the blocks that formed the comet. 
% that PO stems solely from the surface of the comet, nor it being 
%So I assume that also PO survived in the ice whatever the environment was.
%which was stemming from the Southern hemisphere, which was in the winter season at the time. 
% SIMILARITIES between AFGL and comet

%The molecular ratios PO/PN and SO/PN detected in 

We have shown in this work that  PO is more abundant than PN in the massive star-forming region AFGL 5142, confirming previous results in other massive star-forming regions (W51 and W3(OH), \citealt{rivilla2016}). 
%  a protostellar shock, (\citealt{lefloch2016}), and the molecular cloud G+0.693 in the Galactic Center (\citealt{rivilla2018}). 
Interestingly, the analysis of ROSINA data indicate that PO is also more abundant than PN in the comet 67P/C-G by a factor of $>$10. This predominance of PO with respect to PN in both cases is in agreement with the idea that the chemical composition of the comet was incorporated from the natal nebula.
%These results might suggest a chemical link across the full process of star-formation, from the initial parental cloud to the formation of solid bodies in a Solar-like system. 

The ROSINA data also confirm that PO is more abundant than PH$_3$ at least by a factor of 3.3. As explained in Sect. \ref{formation}, the conversion of PH$_3$ to PO needs the action of photochemistry (see Fig. \ref{fig-photochemistry}). Therefore, the ROSINA measurements suggest than the cometary ices suffered some UV processing prior to the formation of the comet, probably due to intense irradiation from neighboring massive stars.

Moreover, we have seen that PO and SO are well correlated in the ROSINA data (Fig. \ref{fig-ROSINA-ratio}), indicating that they are similarly embedded in the cometary ice. In the cavity walls of AFGL 5142, both species are also correlated, spatially and kinematically, suggesting also a similar formation mechanism, as discussed in Section \ref{formation}.  
%The abundance ratio SO/PO in the comet 67P/C-G is $\sim$6 with an uncertainty of a factor of 2. This SO/PO ratio is similar to those detected in some of the P-spots, with values $\sim$10$-$20 in P4, P5 and P6. 
%and lower than that found in P3 and P2 (31 and 80, respectively). 
%However, the SO/PO ratio suffers large variations in AFGL 5142. It is similar to the ratio detected in the comet in some of the P-spots, with values $\sim$10$-$20 in P4, P5 and P6. Towards P3 and P2 the values are higher (31 and 80, respectively).

These similarities may indicate that the chemical composition of the comet, and in particular the PO budget, comes from a shocked (first) and UV-illuminated (afterwards) gas condensation of the massive star-forming cluster that formed the Solar System. In this condensation, after a sudden decrease of the temperature due an increase in density and efficient dust cooling, PO condensed on grains along with other molecules, including complex species also detected in the comet (\citealt{goesmann15,altwegg2017}). Under these physical conditions, deuteration increased, producing heavy water and other deuterated species detected in the comet (\citealt{altwegg2017_deuterium}). PO does not react easily with atomic H, which is the only mobile species on the dust grains at low temperature. This allowed a relatively high abundance of PO in the ices of the comet. 

\subsubsection{From Solar-system bodies to primitive Earth: providing prebiotic P}

The confirmation of the presence of PO in the comet 67P/C-G has important implications for the the prebiotic chemistry that occurred on early Earth.
The possible contribution of prebiotic material from comets is a long-standing topic (e.g. \citealt{chyba1990}). Recent works have pointed out that comets could have contributed significantly to the chemical reservoir of our planet. \citet{Obrien2018} claimed that $\sim$10$\%$ of Earth's water could have its origin in a cometary source, while \citet{marty2017}, based on ROSINA measurements of Xenon isotopes in 67P/C-G, showed that comets might provided the $\sim$22$\%$ of Earth's atmospheric Xe. In sight of our findings of P in 67P/C-G, we explore here the possible contribution of comets to the P reservoir on early Earth.

The amount of P in the Earth crust is 930 ppm (\citealt{yaroshevsky2006}), which means $\sim$2.6$\times$10$^{22}$ g. However, most of this P is not available for biological processes, being locked in insoluble minerals.
%This quantity is several orders of magnitude higher than the amount of P provided by comets. 
%The crust of the Earth has probably his origin in refractory material in rocks, and not in comets.
%As mentioned above, the crust of the Earth has probably his origin in refractory material in rocks.
In the astrobiology community, there is an intense debate (known as the 'phosphate problem', see e.g. \citealt{menor-salvan2018}) about the availability of inorganic phosphate at the dawn of our planet. 
%phosphorus is locked up in certain minerals that life has difficulty making use of.
Orthophosphate (PO$_4^{3-}$), which is the most common form of P on Earth, is mainly found in apatite minerals. The insolubility of these minerals locks P, which is not available for (pre)biotic chemistry. To solve this scarcity of P, alternative sources of P should be explored. 
One possibility comes from meteorites. It has been measured that CI chondrites, a type of stony meteorites, contain a nearly solar P abundance (\citealt{lodders2003}). This suggests that an important fraction of P likely came to the Earth in rocks. However, as mentioned before, most of this P is not easily accesible to trigger prebiotic chemistry due to low solubility in water.  As a possible solution, \citet{pasek2008} proposed that schreibersite, (Fe,Ni)$_3$P, the major carrier of P in iron meteorites, which is more soluble in water, may be the main contributor of P in the early Earth. These authors claimed that that impacts of meteorites probably delivered between 10$^{18}$ and 10$^{21}$ g of P. 

%Pasek et al. (2008) stated that extraterrestrial P would have been a significant component of the total P on the Earth’s surface.

Another extraterrestrial possibility would be the contribution from comets. In this sense, \citet{rubin2019} have recently shown that most of the carbon inventory of our planet may come from comets. Similarly, the recent confirmation of PO, both in star-forming regions and in the comet 67P/C-G, suggests that it is likely that comets provided also an important contribution to the P reservoir on early Earth in the form of PO. Based on data of Xe isotopes (\citealt{marty2017}),  \citet{rubin2019} have deduced that the amount of PO that may have reached the Earth thanks to comets is between (0.2$-$4)$\times$10$^{17}$ g. 
Then, the fraction of PO with respect to the P from meteorites would be in the range 2$\times$10$^{-5}-$0.4. Thus, the amount of P from comet 67P/C-G might be a negligible fraction, or maybe as important as the meteorite contribution. Given the large uncertainties in both estimates, it is difficult to draw a firm conclusion. 

The predominance of PO with respect to PN might have deep implications for prebiotic and biotic chemistry, since the chemical bond between P and O is the basic building block of phosphates, which are essential elements to form large biomolecules such as DNA, RNA, phospholipids and ATP. The large amount of P in star-forming regions, and in particular in the form of an oxygen derivative, PO, implies a high availability of phosphates to be delivered to the early Earth, rather than nitrogen derivatives such as PN. This could explain why prebiotic chemistry seems to prefer PO-based compounds (\citealt{macia1997}), rather than alternative proposed paths for prebiotic chemistry based on nitrogen derivatives (PN-based; \citealt{karki2017}).

\vspace{0.5cm}

%extraterrestrial P would have been a significant component of the total P on the Earth's surface.

%We might be able to propose an alternative scenario, since we have detected significant amount of PO in 67P. Similarly to the carbon inventory, could we state that the P-inventory may also come from comets?

%Since it seems that prebiotic chemistry seems to prefer PO-based compounds rather than PN-based derivatives, this may indicate that the P-bearing reservoir of the early Earth was delivered mainly by comets.

\section{Summary and Conclusions}
\label{conclusions}

To understand the formation of P-bearing molecules in star- and planet-forming regions, we have analysed ALMA high-resolution ($\sim$2$\arcsec$) observations of PN and PO towards the massive star-forming region AFGL 5142, combined with new analysis of the data of the comet 67P/Churyumov-Gerasimenko taken with the ROSINA instrument onboard the {\it Rosetta} spacecraft. 
% SUMMARY ALMA
The ALMA maps have allowed us to study the spatial distribution and kinematics of P-bearing species for the first time in a star-forming region. 
We find that the emission of PN and PO comes from several spots  (P-spots) associated with low-velocity gas with narrow linewidths (1$-$2 km s$^{-1}$) in the cavity walls of the bipolar molecular outflow powered by the central protostar. 
The P-bearing species are likely tracing post-shocked gas in the walls or alternatively slow shocks in the outflow wind, whose
velocity is significantly lower than that of the gas along the outflow axis.
There is no strong emission of P-bearing species towards the central hot molecular core or the starless cold core previously identified in AFGL 5142, nor towards the prominent high-velocity shocks (traced by SO) along the bipolar outtflow axis.
Our Local Thermodynamic Equilibrium (LTE) and non-LTE analyses show that PO is always more abundant than PN in the P-spots  by factors 1.4$-$7.6, with only one exception (P2). We have observed an increasing trend of the PO/PN ratio with distance to the central protostar. We have found that the spatial distribution and the kinematics of PO is more similar to that of SO than that of PN. In particular, PO and PN are tracing gas at velocities shifted by $\sim$0.5 km s$^{-1}$ in all the spots, suggesting that they trace different layers of the cavity walls. 
The interpretation of our ALMA data points towards a scenario for the formation of PO and PN in which shocks are needed to sputter P from the surface of dust grains (in the form of PH$_3$), and photochemistry in the gas phase induced by UV photons from the protostar efficiently forms PO and PN in the illuminated cavity walls. If the gas of the cavity wall then collapses to form, e.g., a Sun-like star, PO can freeze-out and be trapped in the ice mantles until the formation of pebbles, rocks and ultimately comets.
 %SUMMARY ROSINA

The analysis of the ROSINA mass spectroscopy data has revealed a prominent peak at 46.696 Da, which is attributable to PO. PO is indeed the main carrier of P in the comet, with PO/PN$>$10 and PO/PH$_3>$3.3. Similarly to other molecules (O$_2$, S$_2$ or deuterated species), there is evidence that PO was already in the cometary ices prior to the birth of the Sun. In this context, the chemical budget of the comet might have been be inherited from the natal environment of the Sun, which is thought to be a stellar cluster including also massive stars. This scenario is supported by the analysis of P-bearing molecules presented in this work, which has shown chemical similarities between the comet and the AFGL 5142 massive star-forming region, such as the predominance of PO with respect to other P-bearing species, and the good correlation between PO and SO. Finally, the dominant role of PO in the ISM and in the comet might have important implications for the supply of the budget of P on our early Earth, supporting the key role of PO-based molecules (e.g., phosphates) in prebiotic chemistry.

%We have used ALMA to map for the fist time the P-bearing molecule emission (PO and PN) towards a star-forming region, AFGL5141. 
%These observations provide two important results: 1) PO is the main carrier of P in star-forming regions, with PO/PN ratio 5-10; 2) the P-bearing emission arises from the molecular outflow, which clearly indicates that P-bearing molecules can be formed (or released) in shocks that sputter the dust grains. 
%Regarding the 67P/Churyumov-Gerasimenko comet, we have detected P for the first time in a cometary coma with in-situ measurements by ROSINA Spectrometer[9]. At the time of the detection, the parent species could not be determined; however, now a clear signature of PO has been discovered, which is clearly the dominant P-bearing species (upper limits for PN and PH$_3$ have been derived), in remarkably agreement with the results found in AFGL 5142.
%Combining the detection of P-bearing molecules in AFGL 5142 and 67P, it is clear that PO is the dominant source of P, several times more abundant than PN, suggesting a thread between the chemical content of P between a primordial star-forming nebula and Solar System pristine material. This discover has strong implications for the supply of molecular species on our Early Earth, supporting the predominant role of PO-based molecules (e.g. phosphates) in prebiotic chemistry.

\section*{Acknowledgments}
We thank the anonymous referee for her/his instructive comments and suggestions.
This paper makes use of the following  ALMA data:
\noindent
ADS/JAO.ALMA\#2016.1.01071.S
ALMA is a partnership of ESO  (representing its  member states), NSF (USA) and  NINS (Japan), together  with  NRC (Canada), NSC and ASIAA (Taiwan), and KASI (Republic of Korea), in co-operation with the Republic of Chile. The Joint ALMA Observatory is operated by ESO, AUI/NRAO and NAOJ.     
Data from ROSINA, an instrument part of {\it Rosetta} mission, were used in this work. {\it Rosetta} is a European Space Agency (ESA) mission with contributions from its member states and NASA. We acknowledge herewith the work of the entire ESA {\it Rosetta} team over the last 20 years.
This research utilized Queen Mary's MidPlus computational facilities, supported by QMUL Research-IT, http://doi.org/10.5281/zenodo.438045.
This project has received funding from the European Union's Horizon 2020 research and innovation programme under the Marie Sk\l{}odowska-Curie grant agreement No 664931.
MND acknowledges the financial support of the SNSF Ambizione grant 180079, the Center for Space and Habitability (CSH) Fellowship and the IAU Gruber Foundation Fellowship.
The work by AV is supported by the Latvian Council of Science via the project lzp-2018/1-0170.
MR acknowledges the support of the State of Bern, the Swiss National Science Foundation (SNSF, 200021$-$165869 and 200020$-$182418), the Swiss State Secretariat for Education, Research and Innovation (SERI) under contract number 16.0008-2, and the European Space Agency's PRODEX Programme.

\newpage
\bibliographystyle{mnras}
\bibliography{Bib,floris,fhepo,molpro} % if your bibtex file is called example.bib

\appendix

\section{Affiliations}
\label{affiliations}

$^{1}$ INAF-Osservatorio Astrofisico di Arcetri, Largo Enrico Fermi 5, I-50125, Florence, Italy\\
%$^{2}$School of Physics and Astronomy, Queen Mary University of London, Mile End Road, London E1 4NS\\
%$^{3}$Centro de Astrobiolog\'ia (INTA-CSIC). Ctra de Ajalvir, km. 4, Torrej\'on de Ardoz, 28850 Madrid, Spain\\
%$^{3}$University College London, Gower Streer, UK\\
%$^{4}$ESO/European Southern Observatory, Karl Schwarzschild str. 2, D-85748, Garching, Germany\\
%\altaffiltext{5}
%$^{5}$Excellence Cluster “Universe”, Boltzmann str. 2, D-85748 Garching bei Muenchen, Germany\\
%$^{6}$European Southern Observatory, Alonso de Córdova 3107, Vitacura, Santiago, Chile \\
%$^{7}$Joint ALMA Observatory, Alonso de Córdova 3107, Vitacura, Santiago, Chile\\
$^{2}$ Center for Space and Habitability, University of Bern, Gesellschaftsstrasse 6, CH-3012 Bern, Switzerland \\
$^{3}$ Physikalisches Institut, University of Bern, Sidlerstrasse 5, CH-3012 Bern, Switzerland\\
$^{4}$ Max-Planck-Institute for Extraterrestrial Physics, Garching, Germany\\
$^{5}$ SRON Netherlands Institute for Space Research, Landleven 12, 9747 AD Groningen, The Netherlands \\
$^{6}$ Kapteyn Astronomical Institute, University of Groningen, The Netherlands  \\
$^{7}$  Ural Federal University, Ekaterinburg, Russia \\
$^{8}$ Visiting Leading Researcher, Engineering Research Institute 'Ventspils International Radio Astronomy Centre' of Ventspils \\
University of Applied Sciences, In$\check{z}$enieru 101, Ventspils LV-3601, Latvia \\
$^{9}$ LOMC -- UMR 6294, CNRS-Universit\'e du Havre, France  \\
$^{10}$ School of Health, Sport \& Bioscience, University of East London, Stratford Campus, Water Lane, London E15 4LZ, UK \\
$^{11}$ Department of Chemistry and Biochemistry, School of Biological and Chemical Sciences, Queen Mary University of London, Joseph Priestley Building, Mile End Road, London E1 4NS, UK  \\
$^{12}$ESO/European Southern Observatory, Karl Schwarzschild str. 2, D-85748, Garching, Germany\\
%\altaffiltext{5}
$^{13}$Excellence Cluster "Universe", Boltzmann str. 2, D-85748 Garching bei Muenchen, Germany\\
$^{14}$ The ROSINA team: H. Balsiger$^{15}$, J. J. Berthelier$^{16}$, J. De Keyser$^{17}$, B. Fiethe$^{18}$, S. A. Fuselier$^{19}$, S. Gasc$^{15}$, T. I. Gombosi$^{20}$, T. S\'emon$^{15}$, C. -y. Tzou$^{15}$; 
%\vspace{0.5cm}
$^{15}$Physikalisches Institut, University of Bern, Sidlerstrasse 5, CH-3012 Bern, Switzerland. 
$^{16}$LATMOS 4 Avenue de Neptune, F-94100 SAINT-MAUR, France. 
$^{17}$Royal Belgian Institute for Space Aeronomy (BIRA-IASB), Ringlaan 3, B-1180 Brussels, Belgium. 
$^{18}$Institute of Computer and Network Engineering (IDA), TU Braunschweig, Hans-Sommer-Strasse 66, D-38106 Braunschweig, Germany. 
$^{19}$Space Science Division, Southwest Research Institute, 6220 Culebra Road, San Antonio, TX 78228, USA.
$^{20}$Department of Atmospheric, Oceanic and Space Sciences, University of Michigan, 2455 Hayward, Ann Arbor, MI 48109, USA.

\section{Hyperfine excitation of PO by He}
\label{appendix-PO-collision-rates}

Collisional cross sections for the excitation of PO($^2\Pi$) by collisions with He atoms have been computed recently by \citet{lique2018}. In the calculations, the hyperfine structure of the PO radical was taken into account. The $^{31}$P nucleus possesses a non-zero nuclear spin ($I$ = 1/2), which couples with $\vec{j}$ (the rotational momentum) resulting in a splitting of each $\Lambda$-doublet level into two hyperfine levels. The hyperfine levels are labeled by $F$, which is the quantum number
of the total angular momentum $\vec{F} = \vec{j} + \vec{I}$, and takes values between $|j-I|$ and $|j+I|$.

The cross-section calculations were based on a new PO--He potential energy surface (PES) described in detail in \citet{lique2018}. 
Briefly, \textit{ab initio} calculations of the PES of the PO--He van der Waals complex were carried out in the open-shell partially spin-restricted coupled cluster approach at the single, double and perturbative triple excitations [UCCSD(T)] \citep{Hampel:92,watts:93} level of theory using the \texttt{MOLPRO} 2010
package \citep{MOLPRO_brief}. The calculations were performed using the augmented correlation-consistent triple zeta (aug-cc-pVTZ)
basis set \citep{dunning:89} augmented with bond functions defined by \citet{williams:95b}.

Close-coupling quantum scattering calculations were carried out using the HIBRIDON\footnote{The HIBRIDON package (version 4.4) was written by M. H. Alexander,
    D. E. Manolopoulos, H.-J. Werner, B. Follmeg, and P. Dagdigian with
    contributions by D. Lemoine, P. F. Vohralik, G. Corey, R. Gordon,
    B. Johnson, T. Orlikowski, A. Berning, A. Degli-Esposti, C. Rist,
    B. Pouilly, J. K{\l}os, Q. Ma, G. van der Sanden, M. Yang, F. de Weerd,
    S. Gregurick, and F. Lique; {\tt http://www2.chem.umd.edu/groups/alexander/}} program %\citep{hibridon} 
which provided integral cross sections.
Nuclear spin free $S^{J}(F_ij\varepsilon l;F'_ij'\varepsilon' l')$ scattering matrices between
the PO rotational levels were obtained following the standard formalism for collisions of diatomic
open-shell molecules with atoms \citep{alexander:85,kalugina:14}. In the above notation,
$F_i$ denotes the spin-orbit manifold, $l$ the orbital angular momentum quantum
numbers, and $J$ the total angular momentum ($\vec{J} = \vec{j} + \vec{l}$).
The symbols $\varepsilon$, $\varepsilon'$ label the $\Lambda$-doublet level
which can be either $e$ or $f$.
Hyperfine cross sections for transitions from $F_ij\varepsilon F$ to $F_i'j'\varepsilon' F'$ were obtained using a recoupling technique \citep{marinakis:16}. Details on the calculations were provided in \cite{lique2018}.

Here, from the calculated cross sections of \cite{lique2018}, we have obtained the corresponding
thermal rate coefficients at temperature $T$ by averaging over the
collision energy ($E_c$):

\begin{eqnarray}
\label{thermal_average}
k_{\alpha \rightarrow \beta}(T) & = & \left(\frac{8}{\pi\mu k^3_{B} T^3}\right)^{\frac{1}{2}} \nonumber\\
&  & \times  \int_{0}^{\infty} \sigma_{\alpha \rightarrow \beta}\, E_{c}\, e^{-\frac{E_c}{k_{B}T}}\, dE_{c}
\end{eqnarray}

\noindent where $\sigma_{\alpha \rightarrow \beta}$ is the cross section,
$\mu$ is the reduced mass of the system and $k_{B}$ is  Boltzmann's
constant. Calculations up to 960~cm$^{-1}$ allowed determining rate
coefficients from 10 to 150~K. The first 116 hyperfine levels were considered in the calculations.

% Don't change these lines
\bsp	% typesetting comment
\label{lastpage}
\end{document}